\documentclass[journal=macro,manuscript=article]{achemso}

\usepackage[version=3]{mhchem} 

\newcommand*\mycommand[1]{\texttt{\emph{#1}}}

\author{William F. Drayer}
\affiliation[Penn]
{Department of Materials Science and Engineering, University of Pennsylvania,
Philadelphia, PA 19104, USA}
\author{David S. Simmons}
\email{dssimmons@usf.edu}
\affiliation[USF]
{Department of Chemical, Biological and Materials Engineering, University of South Florida, Tampa, FL 33620, USA}

\title[An \textsf{achemso} demo]
{
Is the Molecular Weight Dependence of the Glass Transition Temperature Driven by a Chain End Effect?
}

\abbreviations{IR,NMR,UV}
\keywords{American Chemical Society, \LaTeX}

\begin{document}

\begin{tocentry}

\includegraphics[width=7cm]{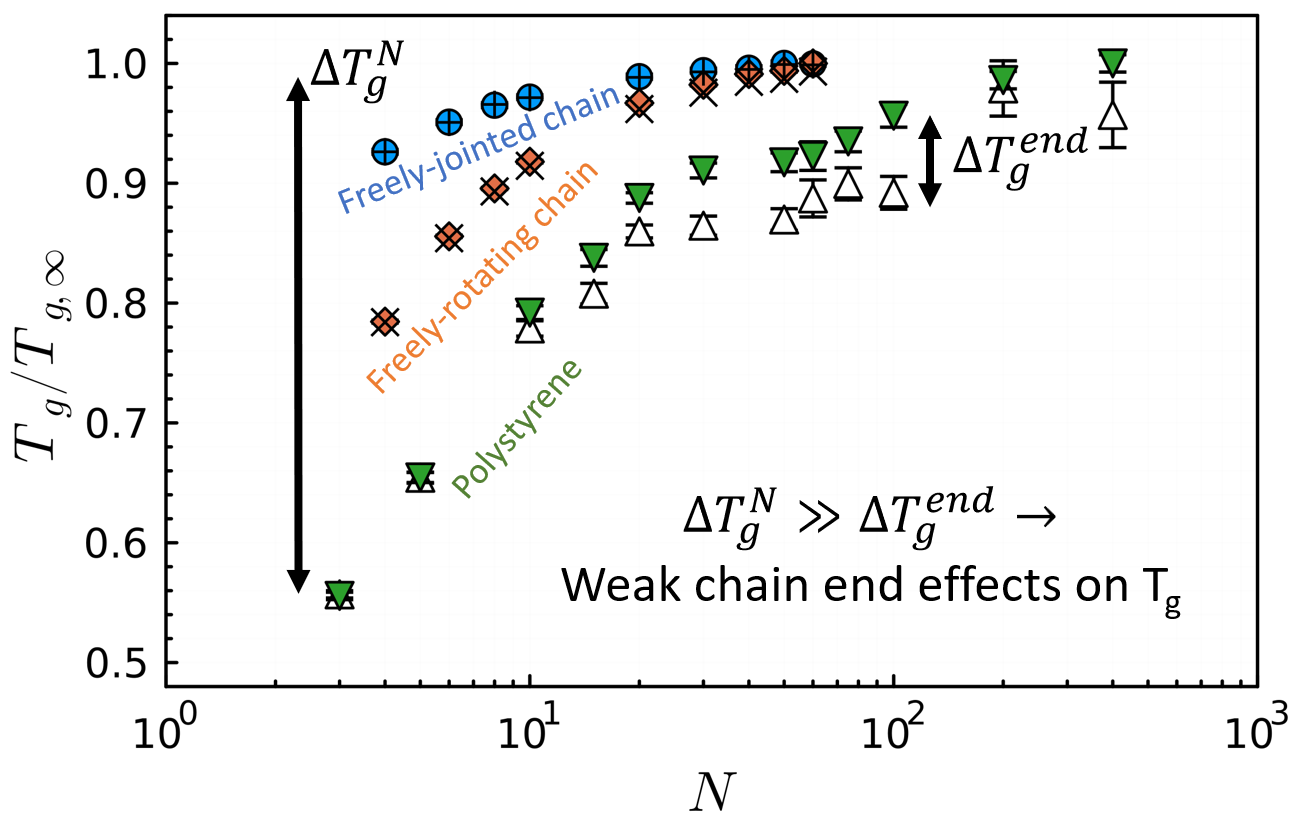}





\end{tocentry}

\begin{abstract}

The immense dependence of the glass transition temperature $T_g$ on molecular weight $M$ is one of the most fundamentally and practically important features of polymer glass formation. 
Here, we report on molecular dynamics simulation of three model linear polymers of substantially different complexity demonstrating that the 70-year-old canonical explanation of this dependence 
(a simple chain end dilution effect)
is likely incorrect at leading order. 
Our data shows that end effects are present only in relatively stiff polymers and,  
furthermore, that the magnitude of this end effect diminishes on cooling.
Instead, we find that $T_g(M)$ trends are instead dominated by shifts in $T_g$ throughout the entire polymer chain rather than through a chain end effect.
We show that these data are consistent with a generic two-barrier model of $T_g$ and its $M$-dependence,
motivated by the Elastically Collective Nonlinear Langevin Equation (ECNLE) theory.
More broadly, this work indicates 
both a need to reassess the canonical understanding of $T_g(M)$ in linear polymers 
(and macromolecules at large) and 
an opportunity to reveal new glass formation physics with renewed study of $M$ effects on $T_g$.


\end{abstract}

\section{Introduction}

A diverse array of systems solidify on laboratory timescales through the glass transition,
a poorly understood phenomenon wherein relaxation times dramatically grow on cooling over a finite range of temperature $T$ \cite{debenedetti_supercooled_2001,cavagna_supercooled_2009,novikov_temperature_2022}.
One central feature of this transition
is a profound dependence on molecular weight $M$ or polymer degree of polymerization $N$: 
the glass transition temperature $T_g$ commonly differs over 200 K 
between the small molecule (e.g., monomer) and the infinite $M$ polymer limits \cite{novikov_correlation_2013}. 
Indeed, 
while this has historically understood as an issue of polymer physics only, 
more recent work has suggested a continuum of $T_g$ size dependence spanning from polymers with large $M$ down to the genuine small molecule limit\cite{novikov_correlation_2013}.

The canonical textbook explanation 
\cite{hiemenz_polymer_2007, 
rubinstein_polymer_2003, 
coleman_fundamentals_1998, 
rudin_elements_2012, 
mathot_calorimetry_1994, 
rosen_fundamental_1993} 
for this trend was established in the early 1950's by Fox and Flory (FF) 
\cite{fox_secondorder_1950} 
and 
Ueberreiter and Kanig (UK) 
\cite{ueberreiter_self-plasticization_1952}.
In the case of the former,
specifically for polystyrene polymers,
end groups are suggested to
``act like a foreign substance in disrupting the local configuration\-al order of the styrene units,''
\cite{fox_secondorder_1950} 
which in turn has been interpreted as a chain end free volume effect present in polymers more generally. 

Ueberreiter and Kanig are more explicit in their interpretation of chain ends and their impact on $T_g$ at large.
The original publication 
\cite{ueberreiter_self-plasticization_1952}
in fact has sections entitled
``Polymers as Mixtures of End and Middle Groups''
and
``Chain End Groups Acting as Plastic\-iz\-ers.''
Their discussion of end groups states that
``[they] have a greater expansion coefficient according to an improved mobility which is due to their privileged position''
and later that ``[i]t therefore seems reasonable to treat the end groups as plastic\-iz\-ers.''
Finally, they remark in their summary that
``[t]he end groups act as plastic\-iz\-ers and cause the `self-plastic\-iz\-a\-tion' of the polymer.''

These arguments are reasonable and intuitively appealing,
with the ends exhibiting some combination of faster dynamics or higher free volume due to having one less bonded neighbor and 
the middle groups exhibiting slower dynamics and/or lower free volume due to their extra bond relative to the end groups.
The effect on $T_g$ for growing chain length is then simply a dilution of this chain end effect just like that of volume.
One would expect chain ends to exhibit enhanced mobility or free volume relative to other chain segments,
and that the infinite molecular weight limit is reached when the majority of segments are beyond the dynamical or structural influence of chain ends and thus exhibit mobility characteristic of an infinite chain. Indeed, given that interactions cannot be infinite range, it \emph{must} follow that any enhancement in mobility by the chain ends must radiate outward from the ends via some gradient along the backbone or through space.
However, this underlying local mechanism has never been fully validated.

Despite the long-standing predominance of this perspective, and
perhaps in part due to the lack of a direct test to date,
questions have emerged over whether this represents the sole, 
or even the dominant, 
mechanism driving the $M$ dependence of $T_g$. 
Early work by Cowie \cite{cown_general_1975} argued for the presence of 
three regimes of $T_g(M)$ behavior ---
a complexity not captured by the two-parameter FF and UK forms.
Novikov and R{\"o}ssler have suggested that the canonical scenario is missing a distinct mechanism that dominates in the low $M$ limit 
\cite{novikov_correlation_2013}. 
Indeed, their work emphasizes a continuity between the molecular weight dependence of $T_g$ for polymers and that for small molecules, with the former merging into the latter in the low molecular weight limit. 
In addition to suggesting the need for an additional mechanism, 
this work therefore also emphasizes the breadth of the importance of understanding how molecular size impacts $T_g$ in both polymers and small molecules. 
Other distinct scenarios have suggested that the $T_g(M)$ dependence is driven 
by the growth of chain stiffness or intra\-molecular activation barriers with $M$,
neither directly mediated by any chain end effect
\cite{mirigian_dynamical_2015,baker_cooperative_2022}. 
Even the basic physical rationale for the FF and UK perspectives has been reconsidered, 
with Zaccone and Terentjev \cite{zaccone_disorder-assisted_2013} 
showing that the the FF equation can be derived by a chain connectivity rather than chain end dilution argument.

Data published by Miwa et al.
are particularly interesting from the perspective of this discussion \cite{miwa_influence_2003}.
They reported that chain ends exhibit a local drop in the temperature of a spin transition
(which they argue is proportional to $T_g$)
for spin-labeled polystyrene, 
but that the chain end spin transition was \textit{itself} $M$-dependent, 
even at fairly high $M$. 
While the enhancement in mobility at chain ends seems in accord with canonical chain end dilution models, 
its local $M$ dependence is not; 
at least to leading order, the polymer is modelled in FF and UK with chain ends that exhibit enhanced mobility or lower $T_g$ for all $M$, 
with a chain of infinite length infinitely diluting this effect wherein almost all polymer segments exhibit $T_g$ of the infinite limit, $T_{g,\infty}$.
The overall $M$ dependence is instead expected to emerge \textit{at the mean chain level} by averaging. 

To assess whether the $T_g(M)$ dependence is predominantly driven by chain end effects as anticipated by the FF and UK models,
we measure local dynamics in molecular dynamics (MD) simulations of three well-established polymers models:
a freely-jointed chain (FJC) \cite{kremer_dynamics_1990,bulacu_molecular-dynamics_2007}, 
a freely-rotating chain (FRC) \cite{bulacu_molecular-dynamics_2007}, 
and OPLS all-atom polystyrene (AAPS) \cite{jorgensen_development_1996,hung_universal_2019,hung_forecasting_2020}. 
These models span a range of complexity and strength of intra\-molecular correlations,
which prior studies 
\cite{sokolov_why_2007,mirigian_dynamical_2015,baker_cooperative_2022,novikov_temperature_2022,zhou_activated_2022} 
have suggested play an important role in $M$ effects on $T_g$.

\section*{Methodology}

\subsection*{Model and Simulation Details}
We study three models spanning from a fully-flexible bead-spring chain to a chemically realistic polymer in this work.
All simulations are performed in LAMMPS \cite{thompson_lammps_2022}.
The simplest model, the freely-jointed chain (FJC), 
uses the standard finite extensible nonlinear elastic (FENE) bond potential
\cite{kremer_dynamics_1990},
$$E = -0.5 K R_0^2  \ln \left[ 1 - \left(\frac{r}{R_0}\right)^2\right] + 4 \epsilon \left[ \left(\frac{\sigma}{r}\right)^{12} - \left(\frac{\sigma}{r}\right)^6 \right] + \epsilon,$$
with particle size $\sigma=1$,
interaction strength $\epsilon=1$,
FENE elastic constant $K=30$, and
maximum bond elongation $R_0 = 1.5$.
To increase chain stiffness, we employ an angle potential to model a freely-rotating chain (FRC),
$$
E = K_\theta [\cos(\theta) - \cos(\theta_0)]^2.
$$
We set the bending constant $K_\theta=25$ and bending equilibrium angle $\cos\theta_0 = -0.333$ (which corresponds to 109.5 degrees), as done in prior work
\cite{bulacu_molecular-dynamics_2007}.
Bead-spring simulations utilize the Stoermer-Verlet time integration algorithm as implemented in LAMMPS with a time\-step of $\tau = 0.005$.
Both the FJC and FRC span chain lengths of $4 \le N \le 60$ beads with total bead counts of $N = 30000$.

To analyze a model with realistic chemical structure, 
we perform additional analysis of OPLS all-atom polystyrene (AAPS) simulations first published in prior work
\cite{hung_universal_2019,hung_forecasting_2020}. 
Full details of those simulations can be found in those prior publications.
Degrees of polymerization for AAPS range from $3 \le N \le 400$ chemical repeat units, 
with a total chemical repeat unit count per simulation of approximately 800
(e.g., there are 160 chains for $N=5$ and two chains for $N=400$).

We utilize the Pre\-SQ simulation protocol \cite{hung_universal_2019,hung_forecasting_2020} 
wherein simulations begin with a high temperature anneal before sequential linear quenches and 
further i\-so\-thermal annealing sufficient to yield equilibrium relaxation times at the mean system level.
These simulations are performed in the i\-so\-thermal-isobaric ($NPT$) ensemble using the Nose-Hoover thermostat and bar\-o\-stat, with both damping parameters set to $\tau=2$.

\subsection{Analysis Details}

Relaxation is determined using the self-part of the intermediate scattering function,

\begin{equation}
\label{eq:fsqt}
F_s(q,t) = \left\langle 
\frac{1}{S} \Sigma^S_k \frac{1}{N} \Sigma^N_j 
\left( \exp{\left( 
-i \mathbf{q} \cdot \left( \mathbf{r}_j(t+s_k) - \mathbf{r}_j(s_k) \right)
\right)} \right)
\right\rangle_{|\mathbf{q}|=q}
\end{equation}
choosing a wave\-number near the first peak in the structure factor:
7.07196/$\sigma_{LJ}$ 
(where $\sigma_{LJ}$ is the Len\-nard-Jones unit of length and is of order 1 nm in real units) 
for both the freely-jointed and freely-rotation chain models and 1.19952/\AA \ for the all-atom polystyrene model. 
We show representative data for $F_s$ as a function of time for each model with a chain length of $N=10$ for all models, 
with the addition of $N=100$ for AAPS,
plotting both the mean system data and chain end data in the SI.
The slow relaxation process within these relaxation functions is then fit to a stretched exponential and 
the alpha relaxation time is defined as the time at which this function decays to 0.2,
as done in several prior works 
\cite{hanakata_local_2012,lang_combined_2014,hung_heterogeneous_2018,hung_universal_2019,hung_forecasting_2020}.

We perform this analysis two ways. 
First, we compute a relaxation time for the entire system by summing in Equation \ref{eq:fsqt} over all segments in the system. 
Second, we perform the analysis for particular repeat unit locations within the chain. 
In the bead spring models, for example, we compute a relaxation time for chain ends by summing over only end beads. 
We also do this for beads bonded to chain ends, beads bonded to those beads, and so on, 
in general computing a mean relaxation time at each position $i$ within the chain, 
where $i=1$ is the chain end, $i=2$ denotes repeat units bonded to a chain end, and so on. 
We perform a similar analysis for AAPS, but in this case for each repeat unit location, 
we average Equation \ref{eq:fsqt} over all atoms within that repeat unit location in all chains.

After computing these relaxation times across a range of temperature, 
$T_g$ is then quantified by fitting these relaxation time $\tau$ and temperature $T$ data to the MYEGA functional form,
suggested by Mauro et al. \cite{mauro_viscosity_2009}, 
which may be written as
\begin{equation}
\label{eq:myega}
\log{\tau} = \log{\tau_\infty} + \frac{A}{T}\exp{\frac{B}{T}}.,
\end{equation}
where $\tau_\infty$, $A$, and $B$ are the fitting parameters.
We rewrite this self-consistently in terms of the glass transition temperature, 
replacing $A$ as a fitting parameter, as
\begin{equation}
\label{eq:myega_tg}
\log{\tau} = \log{\tau_\infty} 
+ \left(\log\tau_{g} - \log{\tau_\infty}\right) 
\frac{T_g}{T} 
\exp{\left(B
\left(\frac{1}{T} - \frac{1}{T_g}\right)
\right)}
\end{equation}
in order to conveniently obtain standard errors on $T_g$ from a least-squares regression upon choosing a timescale $\tau_g$ for $T_g$.
We define $T_g$ at an extrapolated experimental timescale of 100 seconds for AAPS, to allow for experimental comparability
(this comparison has been validated in prior work at the mean system level \cite{hung_forecasting_2020}). 
For the FJC and FRC, we instead employ a computational timescale $T_g$ convention of $\tau_g = 10^{4}$ dimensionless Lennard-Jones time units,
due to the lack of experimental analogue,
as is standard for bead models
\cite{marvin_nanoconfinement_2014,hsu_glass-transition_2016,cheng_design_2018}.

The same process is used for both mean system (cf. Figure 1, filled markers) and 
repeat unit specific (cf. Figure 1, crosses and open markers, and Figure 2, all markers) 
analysis by employing their respective relaxation times as described above; 
we thus obtain $T_g$ at both a whole level and at the level of particular repeat unit positions with the polymer 
(averaged over many chains). 
We note that the definition of a local $\tau$ and thus $T_g$ in this manner has extensive precedent,
particularly from the perspective of interfacial and nano\-confinement effects on dynamics in glass-form\-er\-s
\cite{scheidler_cooperative_2002,
scheidler_relaxation_2003,
scheidler_relaxation_2004,
baschnagel_computer_2005,
peter_thickness-dependent_2006,
kob_non-monotonic_2012,
hocky_crossovers_2014,
kob_nonlinear_2014,
hao_mobility_2021}.
We especially highlight how Priestley and coworkers have reported on the $T_g$ of individual segment locations as a function of number of bonds from a block copolymer junction, which is very similar to our segment-specific $T_g$ reported here\cite{Christie_Register_Priestley_2018}.

We also compute local values of the Debye-Waller factor, 
$\langle u^2 \rangle$,
which is a measure of dynamical free volume 
\cite{mckenzie-smith_relating_2022,
mckenzie-smith_explaining_2021,
hung_universal_2019,
puosi_fast_2019,
pazmino_betancourt_quantitative_2015,
ottochian_universal_2011,
ngai_why_2004,
starr_what_2002,
ngai_correlation_2001},
glassy elasticity 
\cite{van_zanten_brownian_2000,yang_glassy_2011}, and 
particle localization 
\cite{dyre_colloquium:_2006, mirigian_elastically_2014, mirigian_elastically_2014-2, hall_aperiodic_1987, buchenau_relation_1992}
that quantifies the space accessed by a segment within a cage of its transient neighbors. 
Specifically, 
we choose the mean-square displacement (MSD) at a time delta of 0.275 
(or $\approx 10^{-0.55}$) 
dimensionless time units for both the FJC and FRC and 
0.711 ps (or $\approx 10^{-0.15}$ ps) for AAPS (consistent with prior work\cite{hung_universal_2019}).
As $\langle u^2 \rangle$ is a (nearly linear, see SI) function of temperature, 
we choose a consistent $T$ near the lowest accessed by simulation for each model to assess trends with respect to $N$.
The chosen temperatures (0.4 for the FJC, 0.6 for the FRC, and 500 K for AAPS) are at worst
a slight extrapolation for the largest chain lengths
(as the longest chains exhibit the largest $T_g$ values)
with the majority of the data being interpolations.
Plots of the MSD data for select systems and $\langle u^2 \rangle$ as a function of temperature are found in the SI (for the same systems as done for the $F_s$ data).

\section{Results}

\subsection{Quantifying Chain End Effects}

\begin{figure}[t]
 \includegraphics[width=0.5\linewidth]{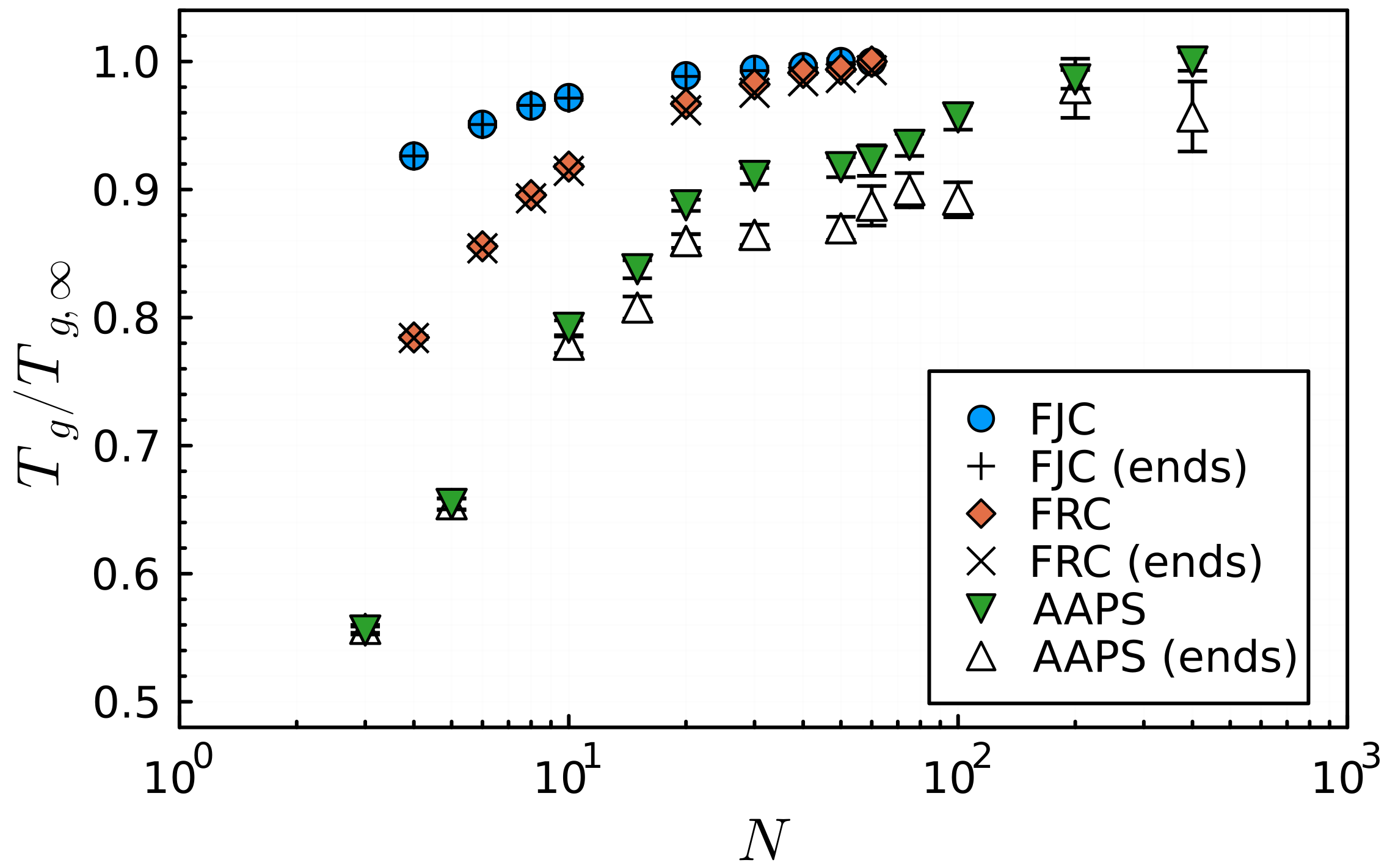}
 \caption{\label{fig:relax_tg_n}
Mean system and chain end $T_g$, normalized by its value for the highest $M$ simulated for a given system, plotted for the systems shown in the legend.
 Error bars are standard errors on $T_g$ as a fit parameter for AAPS (standard errors are smaller than the data points for the FJC and FRC).
 }
\end{figure}

We begin by considering how varying $M$ (by means of degree of polymerization or chain length $N$) impacts $T_g$ for these systems. 
We initially define $T_g$ for AAPS on the experimental timescale (100 seconds) by extrapolating the fit to Equation \ref{eq:myega_tg}. This extrapolation has has been robustly validated against experiment for polystyrene based on this simulation model over a wide range of chain lengths in prior work \cite{hung_forecasting_2020}, such that we can reliably infer experimental-timescale glass formation behavior from these simulations at much shorter times. 
For the two coarse systems, consistent with prior work 
\cite{hsu_glass-transition_2016,xia_molecular_2015, mangalara_tuning_2015,ghanekarade_signature_2023} 
we define $T_g(M)$ on a computational timescale of $10^4 \tau_{LJ}$ (Lennard-Jones time units). 
As shown in Fig.\ \ref{fig:relax_tg_n}, 
all simulated systems exhibit an appreciable dependence of their mean system $T_g$ on $N$, 
in a manner comparable to experiment.
This dependence is stronger in models with stronger intra\-molecular correlations 
(AAPS is the most stiff while the FJC is the most flexible), 
consistent with prior reasoning and common trends in experimental polymers 
\cite{sokolov_why_2007,novikov_correlation_2013,mirigian_dynamical_2015,novikov_temperature_2022,baker_cooperative_2022}.

\begin{figure*}[t]
 \includegraphics[width=\linewidth]{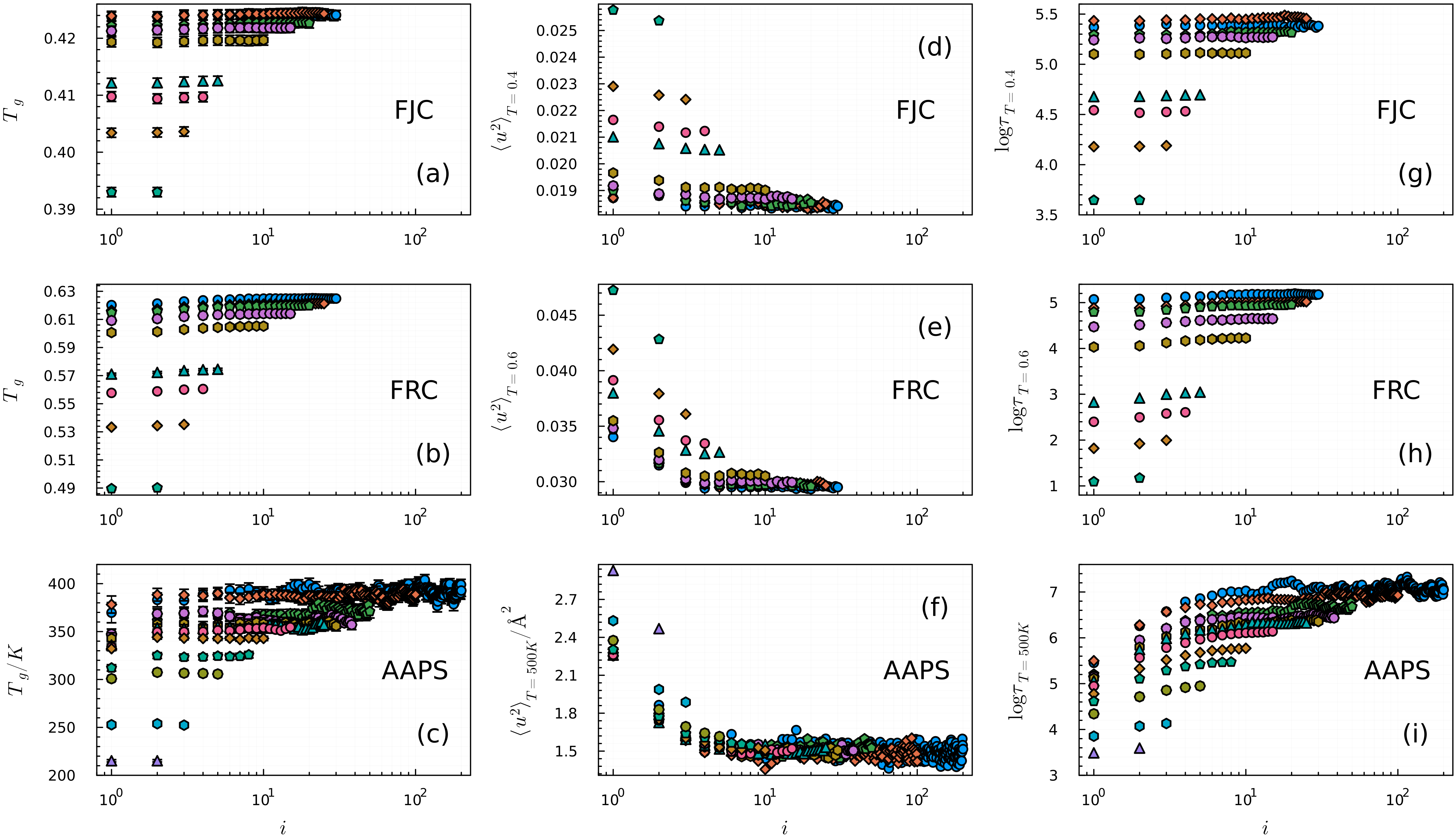}
 \caption{\label{fig:chain_index}
 $T_g$ (a-c), $\langle u^2 \rangle$ (d-f), and $\log \tau$ (g-i) plotted as a function of bead index $i$: 
 chain ends are labeled $i=1$ (either bead for FJC and FRC or monomer for AAPS), 
 $i=2$ indicates repeat units bonded to chain ends, 
 and so on until $i=N/2$, which indicates pairs of middle-most repeat units for even $N$ 
 (for odd AAPS chain lengths, the last data point is a single monomer).
 Each row corresponds to each model, as labeled inside each panel.
 Error bars are standard errors on $T_g$ as a fit parameter (within the data points for FRC).
 Note that error bars increase with chain length for AAPS due to reduced statistical sampling.
 }
\end{figure*}

Additionally shown in Fig.\ \ref{fig:relax_tg_n}, 
all three models exhibit negligible or weak local reduction in $T_g$ at chain ends, 
relative to the overall magnitude of the trend of mean $T_g$ with $N$. 
Essentially no $T_g$ end effect is observed in the FJC model and there is less than a 1\% end reduction for the FRC model 
($\sim 4$ K in real units for typical glassy polymers). 
In AAPS the end effect is of order 30K, 
relative to a shift of nearly 200K in mean $T_g$ with $M$ 
(consistent with Miwa et al.'s experimental findings \cite{miwa_influence_2003}). 
We expand upon this in Figure \ref{fig:chain_index}a-c,
which shows that local variations in $T_g$ along the backbone near the chain end are weak or absent in all three models.
Even in AAPS, where a modest $T_g$ end effect is present, it does not appreciably propagate to covalent\-ly connected segments. 
We instead observe a whole-chain effect of increasing $T_g$ with respect to $N$ regardless of $i$,
in contrast to the expected chain-end effect.
Evidently, 
a direct $T_g$ end effect is vastly too small to account for the much larger variation in mean-chain $T_g$ with $N$ observed in Fig.\ \ref{fig:relax_tg_n}
for all three models.

Could a more pronounced end effect still be present in free volume, which is the underlying proposition of the FF model?
Interestingly, Fig.\ \ref{fig:chain_index}d-f indicate that trends in $\langle u^2 \rangle$ with $M$ and chain location are non\-universal. 
As with $T_g$, shifts in $\langle u^2 \rangle$ with $M$ in the FJC model are nearly uniform through the whole chain, 
with at most a very weak chain end gradient. 
In contrast, AAPS exhibits a significant chain end effect of enhanced $\langle u^2 \rangle$. 
The FRC model exhibits a mix of these two scenarios. 
This trend is perhaps as expected: 
intra\-molecular correlations 
contribute to the cage scale in stiffer polymers, 
such that the reduced bond connectivity near chain ends relieves caging more significantly in these systems. 
However, as shown by the FJC, 
it is evidently possible for a polymer to exhibit at least a 10\% drop in $T_g$ without a significant chain end effect 
in either $T_g$ or $\langle u^2 \rangle$.
This suggests that chain end mobility or free volume effects cannot be the \textit{sole} driver of the $T_g(M)$ dependence. 
In contrast, data for FRC and AAPS indicate that even the presence of an appreciable
enhancement in chain-end $\langle u^2 \rangle$ does not necessarily translate to a substantial suppression in chain end $T_g$.

\begin{figure}[!ht]
 \includegraphics[width=0.5\linewidth]{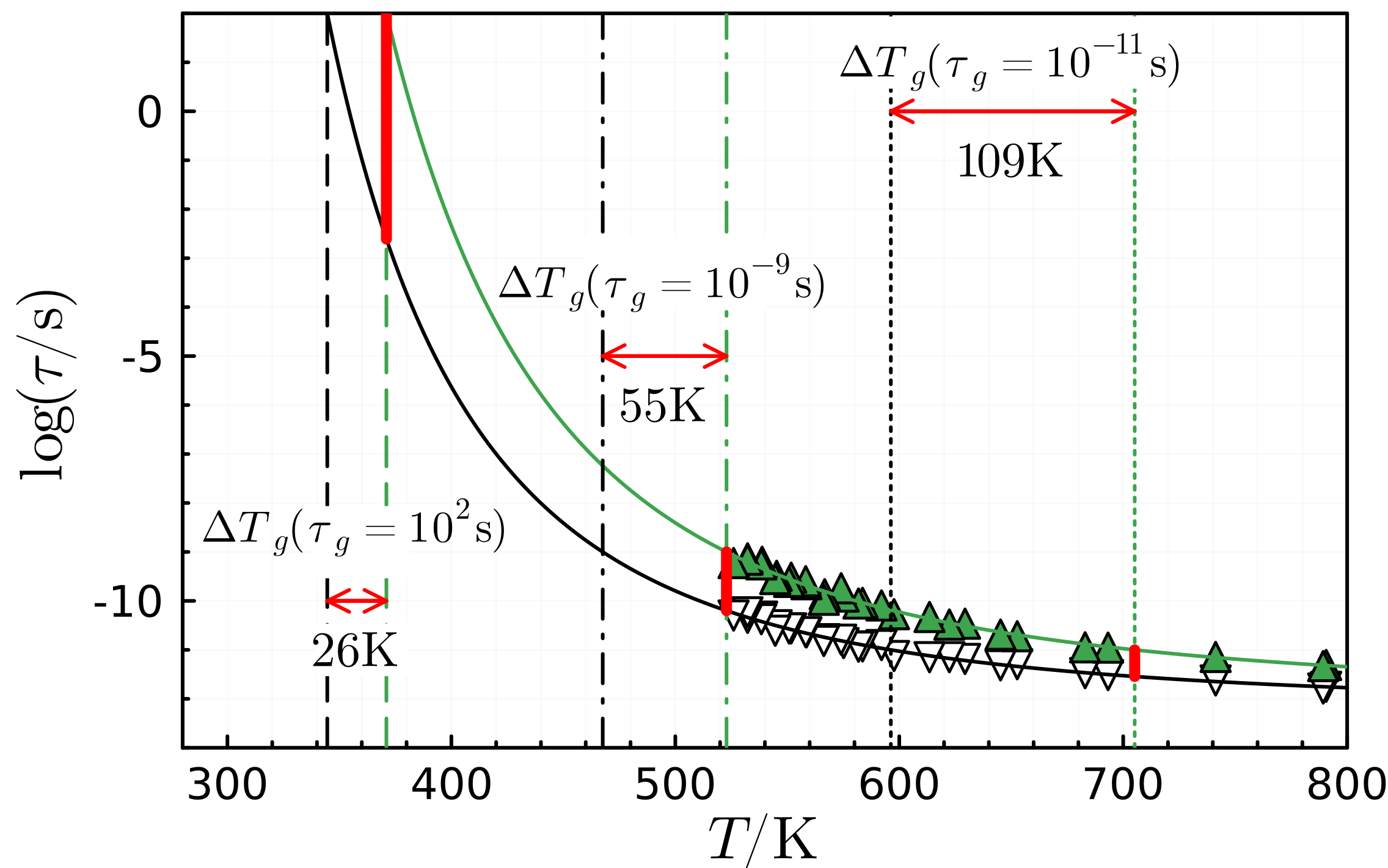}
 \caption{\label{fig:extrapolate}
 $\tau$ vs $T$ for $N=100$ AAPS, for chain ends (open symbols) and mean system (green symbols).
 Curves in corresponding colors are fits to the MYEGA functional form \cite{mauro_viscosity_2009}.
 Green vertical lines (rightmost of each pair) denote the $T$ at which the mean system $\tau$ equals $10^{-11}$s (dotted), 
 $10^{-9}$s (dot-dashed), and $10^2$s (dashed), respectively. 
 Black vertical lines (leftmost of each pair) denote the $T$ at which the chain ends exhibit these same $\tau$.
 The $T$ difference, $\Delta T_g$, is reported for each of the three timescales.
 Heavy red vertical segments highlight the chain end $\tau$ reduction
 relative to the mean system at each timescale. 
 }
\end{figure}

\begin{figure}[!b]
 \includegraphics[width=0.5\linewidth]{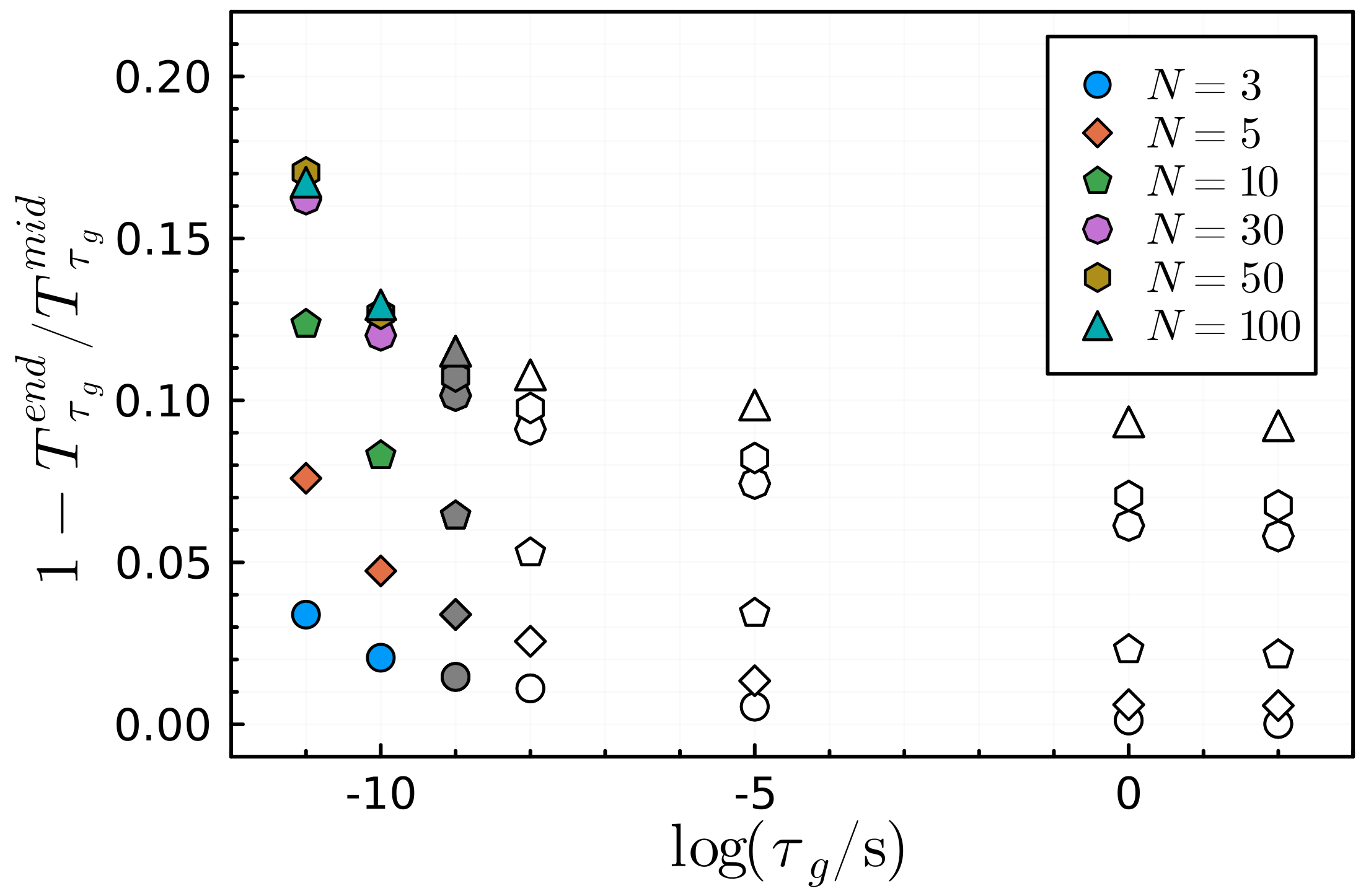}
 \caption{\label{fig:timescale}
 Magnitude of chain end $T_g$ effect vs conventional timescale $\tau_g$ for AAPS chain lengths indicated in the legend.
 Filled markers denote points for which both mid-chain and end-chain $T_g$ values are interpolated from simulation data; 
 grey markers are points for which the mid-chain $T_g$ is interpolated and end-chain values are mildly extrapolated; 
 open markers denote points for which $T_g$ values are extrapolated for all monomers.
 }
\end{figure}

How does enhanced chain end $\langle u^2 \rangle$ not directly lead to suppressed chain end $T_g$? 
In Fig.\ \ref{fig:chain_index}g-i we report relaxation time $\tau$ gradients 
along the chain backbone at the same $T$ for which we reported $\langle u^2 \rangle$ end gradients.
At least for AAPS, 
the chain end $\tau$ is indeed considerably reduced for $T$ well above that of $T_g$ when defined on the experimental timescale.
The failure of this increase to lead to a chain-end $T_g$ enhancement lies in a subtle feature of its $T$ dependence,
which we demonstrate in Fig.\ \ref{fig:extrapolate}.
The chain-end mobility enhancement strengthens on cooling on an \emph{absolute} basis;
however, this strengthening is insufficient to keep up with the growth in the overall activation barrier of relaxation on cooling, 
such that it becomes \emph{relatively} weaker in its implications for $T_g$. 
We further show this effect in Fig. \ref{fig:timescale},
which emphasizes that this effect occurs within computationally accessible timescales
and is not merely a result of extrapolation.
Figures \ref{fig:extrapolate} and \ref{fig:timescale} demonstrate how the magnitude of the $T_g$ end effect shrinks as the timescale $\tau_g$ that defines $T_g$ is increased.

\subsection{Theoretical Interpretations}

Our data indicate that the chain end mobility effects intuited by FF and UK are present at high $T$ in semi\-flexible chains 
(although negligible in quite flexible chains) but diminish in significance upon cooling and become sub-dominant by the experimental $T_g$ timescale. 
Within the context of many classical theories of glass formation, this observation seems surprising. 
Many of these theories, 
including free volume theory \cite{doolittle_studies_1951,white_explaining_2017} and 
classical entropy theories \cite{adam_temperature_1965},
postulate the presence of a single dominant activation barrier to relaxation in glass-forming liquids. 
This barrier is postulated to grow on cooling in a manner that results from a \emph{multiplicative product} of the high temperature barrier with a temperature-dependence amplification factor associated with a growing cooperative lengthscale. 
As an example, within the Adam-Gibbs theory of glass formation, 
$\log(\tau)$ goes as a high-$T$ barrier times a co\-op\-er\-a\-tiv\-i\-ty  factor over $k_BT$ \cite{adam_temperature_1965}. 
A reduction in either high-$T$ activation barrier or co\-op\-er\-a\-tiv\-i\-ty at the chain ends would thus not be expected to diminish in importance on cooling. 
A similar intuition would seem to hold for free volume approaches given the inverse proportionality of the activation barrier to a single quantity (the free volume). 
It is this intuition, that the alteration in activation barriers near the chain end becomes highly important in the glass formation range, 
that drives the classical FF and UK viewpoints.

More recently, a distinct alternative perspective has emerged that views glassy super-Arrhenius behavior as emerging from an \emph{additive}, 
rather than multiplicative, growth in the barrier on cooling 
\cite{mirigian_elastically_2014,mirigian_elastically_2014-2, schmidtke_temperature_2015}. 
In particular, 
the Elastically Collective Nonlinear Langevin Equation (ECNLE) theory of glass formation 
formulates its activation barrier as a sum of a local barrier 
(which grows relatively weakly on cooling) and a collective elastic barrier 
(which is predicted to emerge and then grow relatively strongly on cooling towards $T_g$) 
\cite{mirigian_elastically_2014,mirigian_elastically_2014-2}.

We can understand our results within the context these newer perspectives. 
Consider a generic two-barrier model 
wherein the total activation barrier $F_{tot}^{mid}$ in the mid-chain is a sum 
of a local barrier $F_{loc}^{mid}$ and a collective barrier $F_{coll}^{mid}$:
\begin{equation}\label{eq:midF}
F_{tot}^{mid}\left( N,T \right)=F_{loc}^{mid}\left( N,T \right)+F_{coll}^{mid}\left( N,T \right).
\end{equation}
Alterations of this barrier at the chain end are rooted in alterations to intra\-molecular correlations 
that are intrinsically present arbitrarily far above $T_g$, 
and therefore most directly impact the local barrier 
(since the collective barrier is absent far above $T_g$). 
We thus model the end effect as a fractional reduction of the local barrier by a factor $\alpha^{end}$, which we model as roughly temperature-invariant 
(but expect to be chemistry dependent and larger for stiffer chains) 
since it reflects a truncation of intra\-molecular barriers that are mainly ster\-ic and bonding in nature and therefore relatively a\-thermal:
\begin{equation}\label{eq:endF}
F_{tot}^{end}\left( N,T \right)={{\alpha }^{end}}F_{loc}^{mid}\left( N,T \right)+F_{coll}^{mid}\left( N,T \right).
\end{equation}

One can quantify the expected temperature dependence of the chain-end relaxation time gradient within this perspective by employing the total barrier forms above within a generalized activation law to compute the ratio of chain-end to chain-mid relaxation times:
\begin{equation}\label{eq:tauratio}
\log \left( \frac{{{\tau }_{end}}\left( N,T \right)}{{{\tau }_{mid}}\left( N,T \right)} \right)=\frac{\left( 1-{{\alpha }^{end}} \right){{F}_{loc}^{mid}}\left( N,T \right)}{kT}.
\end{equation}
This equation anticipates that the enhanced chain end mobility (relative to the mid-chain) should grow on cooling, 
as seen above in Fig. \ref{fig:extrapolate} (the vertical red bars, moving from right to left), 
simply because of the reduction in temperature and any growth on cooling of $F_{loc}^{mid}$.

We can further combine Equations \ref{eq:midF} and \ref{eq:endF} with a generalized activation law as above, rearrange them to solve for temperature, apply each at its corresponding local $T_g$,
and take their ratio. This gives
\begin{equation}\label{eq:Tgrad}
\frac{T_{g}^{end}}{T_{g}^{mid}}=R\left[ 1-\left( 1-{{\alpha }^{end}} \right)x_{loc}^{mid}\left( N,T_{g}^{end} \right) \right],
\end{equation}
in which we define two dimensionless ratios: a pre\-factor 
$R={F_{tot}^{mid}\left( N,T_{g}^{end} \right)}/{F_{tot}^{mid}\left( N,T_{g}^{mid} \right)}$ 
and
$x_{loc}^{mid}\left( N,T \right),$
which is the fraction of the total barrier in the mid-chain that is contributed by the local barrier at $T$ such that
$x_{loc}^{mid}\left( N,T_{g}^{end} \right) = {F_{loc}^{mid}\left( N,T_{g}^{end} \right)}/{F_{tot}^{mid}\left( N,T_{g}^{end} \right)}\;$

For a weak $T_{g}$ end effect,
we can approximate the relaxation process as Arrhenius over the limited temperature range involved, giving $R\approx1$. 
Equation \ref{eq:Tgrad} then predicts that the $T_g$ end effect shrinks on cooling, 
even as the $\tau$ end effect grows, 
because the fractional reduction from the local barrier $x_{loc}^{mid}$ shrinks on cooling 
such that the end effects become diluted within the faster-growing overall barrier.
Indeed, this is the behavior we see in Fig. \ref{fig:timescale},
because the fractional reduction from the local barrier $x_{loc}^{mid}$ shrinks on cooling \cite{mirigian_elastically_2014} 
such that the end effects become diluted within the faster-growing overall barrier.

\subsection{What causes $T_g(M)$ if not chain end effects?}

Across three models, 
chain ends evidently do not exhibit sufficiently reduced $T_g$ values to account for the dependence of mean $T_g$ on chain length 
on the basis of a chain end dilution effect. 
This would appear to demand a reevaluation of the textbook understanding of the $T_g(M)$ dependence. 
Our data indicate that $T_g(M)$ variations are primarily driven by the $M$ dependence of the mid-chain (or whole-chain) activation barrier 
$F_{tot}^{mid}(N,T)$, 
which in turn may result from whole-chain trends in some combination of the local and collective barriers. 
This type of scenario has been predicted within the ECNLE theory, 
where both local and collective elastic contributions to this activation barrier grow with increasing size of the fundamental dynamical repeat unit. 
This unit is taken to be the entire molecule in small rigid molecules 
\cite{mirigian_elastically_2014,mirigian_elastically_2014-2} 
and as the Kuhn segment in polymers \cite{mirigian_dynamical_2015,zhou_activated_2022}. 
In the small molecule case, this leads to the prediction that $T_g\sim \sqrt{M}$ \cite{mirigian_elastically_2014,mirigian_elastically_2014-2}, 
which is consistent with the small molecule limit identified by Novikov and R\"ossler \cite{novikov_correlation_2013}. 
In the polymer case, the $M$ dependence follows from the growth of the Kuhn segment with increasing $N$ 
(as measured by growth of the chain's characteristic ratio $C_N$):
a reflection of increases in effective chain stiffness with $M$. \cite{mirigian_dynamical_2015,zhou_activated_2022}. 

\begin{figure}[bt!]
 \includegraphics[width=0.5\linewidth]{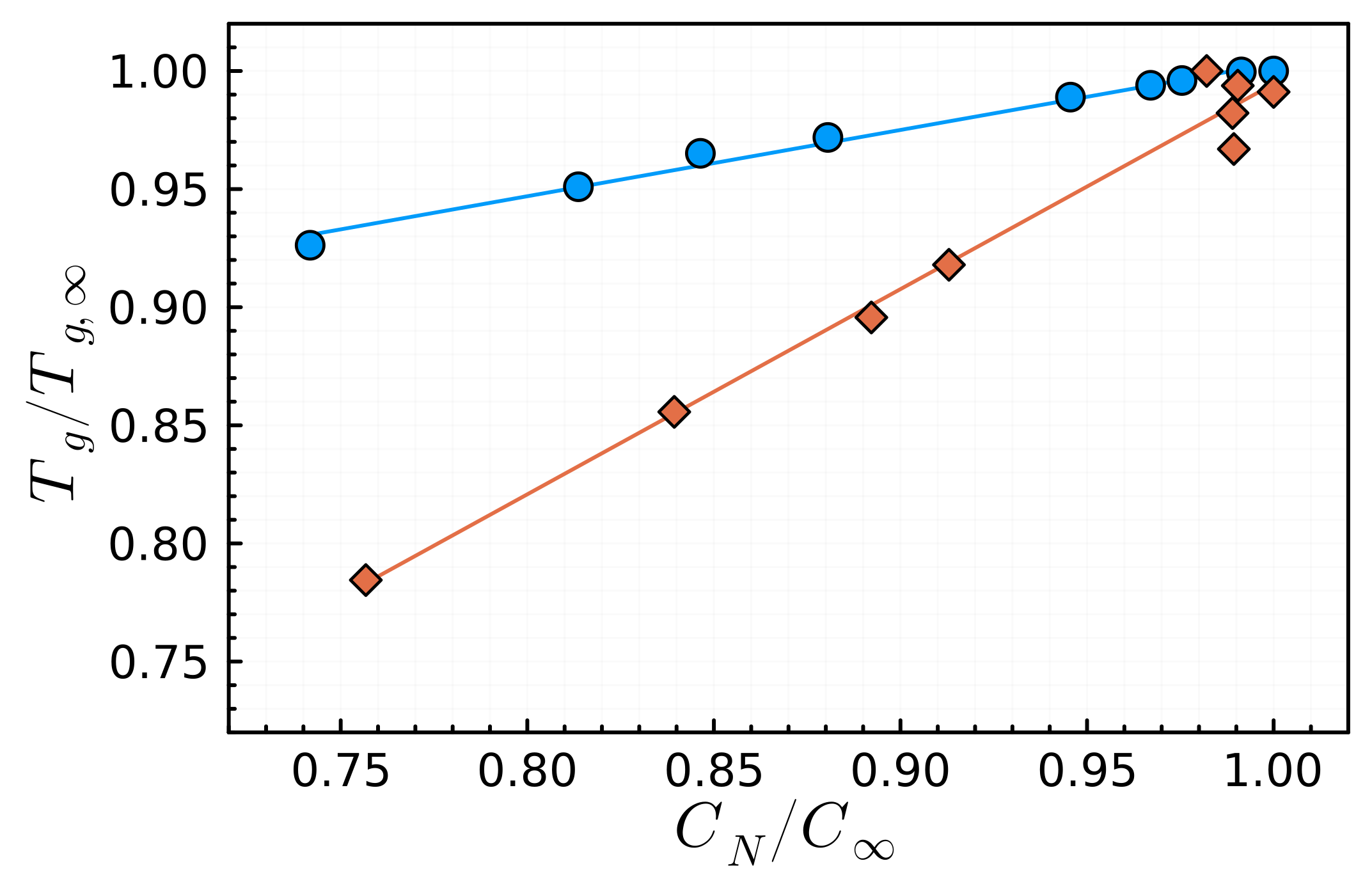}
 \caption{\label{fig:cn}
$T_g$ plotted as a function of normalized $C_N$, each normalized by their value for the longest chain of that type simulated,
 for the FJC (blue circles) and FRC (orange diamonds). Lines in corresponding colors are linear fits.
 }
\end{figure}

Indeed, prior studies have found that variation of $T_g$ with $M$ tracks with variations in $C_N$ 
for at least polystyrene, poly(methyl meth\-acrylate), and polyethylene. 
This has also been reported in poly\-di\-methyl sil\-ox\-ane, although this is in dispute \cite{baker_cooperative_2022}. 
While we cannot add to the extant data for PS $T_g$ vs $C_N$ correlations 
due to the limited number of chains that can be simulated in an AA simulation accessing the glass formation range, 
Fig.\ \ref{fig:cn} illustrates that $T_g$ is proportional to $C_N$ for our two bead-spring systems, 
adding to evidence that $T_g(M)$ closely tracks with $C_N(M)$ as $M$ is varied for a given polymer, 
in line with the ECNLE scenario  \cite{mirigian_dynamical_2015,zhou_activated_2022}.

It is not clear whether this $C_N$ scenario alone fully accounts for the $T_g(M)$ dependence given suggestions that $T_g(M)$ exhibits multiple regimes, 
particularly in stiffer polymers. 
Baker et al. have argued that this may result from nontrivial variations in chain conformation\-al statistics, 
combined with an intra\-molecular dynamical facilitation effect \cite{baker_cooperative_2022}. 
While not excluding that scenario, 
our results suggest a potential alternate scenario given that we find chain end effects on $T_g$ to be present, 
if weak, in our stiffer systems. 
It may be that the multiple regimes observed in some stiffer polymers reflect a combination of a leading order stiffness effect with a perhaps second-order end effect with parallels to FF and UK.

\section{Conclusions}

We have reported on local dynamics for three model polymers of varying degrees of complexity and stiffness, 
ranging from a freely-jointed polymer chain,
with only excluded volume added relative to a random walk in three dimensions,
to a fully atom\-ist\-ic polystyrene chain which indeed agrees with experimental $T_g(N)$ quite well
\cite{hung_forecasting_2020}. 
Surprisingly, 
we find that chain-end mobility enhancements in dynamics or free volume 
are insufficient to account for the shift in overall system $T_g$ with varying M. 
Indeed, the freely-jointed-chain model exhibits an almost complete absence of chain end gradient in $T_g$, 
relaxation time, or Debye-Waller factor $\langle u^2 \rangle$ (one measure of a dynamic free volume). 
Nevertheless, this model exhibits a 10\% reduction in $T_g$ from the highest to lowest molecular weight simulated.

The results for the freely-rotating-chain model and all-atom polystyrene are more complex, 
exhibiting an enhancement in $\langle u^2 \rangle$ at the chain end. However, as noted above, 
this does not translate into  appreciable $T_g$ gradients at chain ends. 
Within many classical theories of the glass transition, 
this would be difficult to understand. 
However, we show that it can be understood in terms of a simple two-barrier scenario of the glass transition inspired by the 
Elastically Collective Nonlinear Langevin Equation (ECNLE) theory of glass formation
\cite{mirigian_elastically_2014,mirigian_elastically_2014-2}, 
which predicts additive contributions to the activation barrier for relaxation 
from a local barrier and an longer-ranged collective barrier that grows more strongly on cooling. 
If chain-end enhancements in $\langle u^2 \rangle$ primarily alter the local barrier contribution, 
we show that the they become less important at low temperatures where the collective barrier dominates.  

Does the absence of a locally-driven chain-end $T_g$ suppression sufficiently large to account for the $T_g(M)$ dependence via spatial averaging 
genuinely provide compelling evidence that some other mechanism must be dominant? 
As we discuss in the introduction, 
any model in which faster dynamics or lower $T_g$ are effectively \emph{nucleated} by a chain end effect 
requires that this effect must emanate over a \emph{finite} distance from the chain end, 
since correlations in super\-cooled liquids are not of infinite range. 
Some gradient in these properties is therefore to be expected. 
Our longest chains are 400 repeat units in the case of AAPS. 
Most segments in this system are not within the first neighbor shell of an end; 
at this molecular weight, chain ends are expected to be approximately 6 segmental diameters apart on average. 
We nevertheless do not observe gradients remotely large enough to account for $T_g(M)$. 

Could a gradient emanating from chain ends simply be so long-ranged as to appear flat even over this spacing? 
There are several reasons to conclude that this is not the case.
First, prior simulations probing the range of correlation length scales in glass-forming liquids 
have suggested that they are of range only a few segmental diameters on the timescales we access; 
they should therefore be readily observable within Figure \ref{fig:chain_index} were they present. 
As a possibly even stronger argument, 
let us consider what it would imply regarding the behavior of even longer chains 
if some dramatically longer range were in fact concealing the gradient within our simulations. 
This would require that a $T_g$ gradient of magnitude comparable to the overall $T_g(M)$ trend 
would have to emerge in Figure \ref{fig:chain_index} for PS chains considerably longer than 400 repeat units (about 40,000 g/mol). 
Since it would be incoherent for the $T_g$ at any location in the chain to drop with increasing molecular weight, 
this would imply that the high-molecular weight mid-chain $T_g$ would have to grow by many tens of K beyond that observed here. 
This is not plausible, 
since the mean $T_g$ value here has already plateaued in accord with high-molecular weight $T_g$ observed for PS in experiment.

Collectively, these findings suggest that chain end effects are not the dominant origin of the $T_g(M)$ dependence. 
They are entirely absent in at least one system exhibiting a substantial $T_g(M)$ dependence, 
and even when present, their effects grow weaker on cooling and play little role in $T_g$ at experimental timescales. 
Indeed, we note that even the modest chain end $T_g$ suppression experimentally inferred by Miwa et al. \cite{miwa_influence_2003} 
was based on a shorter-timescale spin transition that occurs at higher temperatures 
where this analysis would expect chain end effects to be modestly more important in some systems than they are at experimental timescales. 
Perhaps the most remarkable conclusion is that for the fully flexible chain,
chain end effects for \emph{both} $\tau$ and $\langle u^2 \rangle$ are negligible,
from which one must conclude that \emph{at least} another mechanism must be responsible for the observed 10\% reduction in $T_g$ for this model. 
Consistent with prior work, 
we report that $T_g$ in our freely-jointed and freely-rotating chain models 
tracks linearly with the characteristic ratio is the molecular weight is varied. 
This may be consistent with a scenario encoded within the ECNLE theory 
wherein $T_g(M)$ is driven by stiffness variations with molecular weight, 
as indicated by the growth in the characteristic ratio in longer chains. 
More broadly, this may align with 
the proposition that intra\-molecular activation barriers, 
which play a central role in polymer glass formation \cite{colmenero_are_2015}, 
may qualitatively vary with molecular weight \cite{baker_cooperative_2022}.

Overall, these findings indicate a need to reopen the study of $M$ effects on $T_g$, 
with a focus on more recent theories wherein this trend is dominated by whole-chain effects rather than end effects. 
Indeed, the finding that the $T$-dependence of dynamical chain end effects can be understood based on a two-barrier model of the glass transition 
suggests that a renewed focus on this problem may have the potential to yield broader insights into the nature of glass formation itself.

\begin{acknowledgement}

This material is based upon work supported by the National Science Foundation under Grant No. DMR - 1849594. 
The authors acknowledge helpful discussions with Dr. Kenneth Schweizer.



\end{acknowledgement}

\begin{suppinfo}

The supporting information contains curves for decay of the self-part of the intermediate scattering function, the time evolution of the mean-square displacement and the temperature dependence of the Debye-Waller factor for representative chains of each model, along with temperature dependence of mean relaxation time for each model and all chain length.

\end{suppinfo}

\bibliography{achemso-demo}

@PREAMBLE{
 "\providecommand{\noopsort}[1]{}" 
 # "\providecommand{\singleletter}[1]{#1}%" 
}

@article{baschnagel_computer_2005,
	title = {Computer simulations of supercooled polymer melts in the bulk and in confined geometry},
	volume = {17},
	issn = {0953-8984},
	url = {https://dx.doi.org/10.1088/0953-8984/17/32/R02},
	doi = {10.1088/0953-8984/17/32/R02},
	pages = {R851},
	number = {32},
	journaltitle = {Journal of Physics: Condensed Matter},
	shortjournal = {J. Phys.: Condens. Matter},
	author = {Baschnagel, J. and Varnik, F.},
	urldate = {2023-12-14},
	date = {2005-07},
	langid = {english},
}

@article{kob_nonlinear_2014,
	title = {Nonlinear dynamic response of glass-forming liquids to random pinning},
	volume = {90},
	url = {https://link.aps.org/doi/10.1103/PhysRevE.90.052305},
	doi = {10.1103/PhysRevE.90.052305},
	pages = {052305},
	number = {5},
	journaltitle = {Physical Review E},
	shortjournal = {Phys. Rev. E},
	author = {Kob, Walter and Coslovich, Daniele},
	urldate = {2023-12-14},
	date = {2014-11-19},
	note = {Publisher: American Physical Society},
}

@article{hocky_crossovers_2014,
	title = {Crossovers in the dynamics of supercooled liquids probed by an amorphous wall},
	volume = {89},
	url = {https://link.aps.org/doi/10.1103/PhysRevE.89.052311},
	doi = {10.1103/PhysRevE.89.052311},
	pages = {052311},
	number = {5},
	journaltitle = {Physical Review E},
	shortjournal = {Phys. Rev. E},
	author = {Hocky, Glen M. and Berthier, Ludovic and Kob, Walter and Reichman, David R.},
	urldate = {2023-12-14},
	date = {2014-05-23},
	note = {Publisher: American Physical Society},
}

@article{kob_non-monotonic_2012,
	title = {Non-monotonic temperature evolution of dynamic correlations in glass-forming liquids},
	volume = {8},
	rights = {2011 Springer Nature Limited},
	issn = {1745-2481},
	url = {https://www.nature.com/articles/nphys2133},
	doi = {10.1038/nphys2133},
	pages = {164--167},
	number = {2},
	journaltitle = {Nature Physics},
	shortjournal = {Nature Phys},
	author = {Kob, Walter and Roldán-Vargas, Sándalo and Berthier, Ludovic},
	urldate = {2023-12-14},
	date = {2012-02},
	langid = {english},
	note = {Number: 2
Publisher: Nature Publishing Group},
}

@article{scheidler_relaxation_2004,
	title = {The Relaxation Dynamics of a Supercooled Liquid Confined by Rough Walls},
	volume = {108},
	issn = {1520-6106},
	url = {https://doi.org/10.1021/jp036593s},
	doi = {10.1021/jp036593s},
	pages = {6673--6686},
	number = {21},
	journaltitle = {The Journal of Physical Chemistry B},
	shortjournal = {J. Phys. Chem. B},
	author = {Scheidler, Peter and Kob, Walter and Binder, Kurt},
	urldate = {2023-12-14},
	date = {2004-05-01},
	note = {Publisher: American Chemical Society},
}

@article{scheidler_relaxation_2003,
	title = {The relaxation dynamics of a confined glassy simple liquid},
	volume = {12},
	issn = {1292-895X},
	url = {https://doi.org/10.1140/epje/i2003-10041-7},
	doi = {10.1140/epje/i2003-10041-7},
	pages = {5--9},
	number = {1},
	journaltitle = {The European Physical Journal E},
	shortjournal = {Eur. Phys. J. E},
	author = {Scheidler, P. and Kob, W. and Binder, K.},
	urldate = {2023-12-14},
	date = {2003-09-01},
	langid = {english},
}

@article{scheidler_cooperative_2002,
	title = {Cooperative motion and growing length scales in supercooled confined liquids},
	volume = {59},
	issn = {0295-5075},
	url = {https://iopscience.iop.org/article/10.1209/epl/i2002-00182-9/meta},
	doi = {10.1209/epl/i2002-00182-9},
	pages = {701},
	number = {5},
	journaltitle = {Europhysics Letters},
	shortjournal = {{EPL}},
	author = {Scheidler, P. and Kob, W. and Binder, K.},
	urldate = {2023-12-14},
	date = {2002-09-01},
	langid = {english},
	note = {Publisher: {IOP} Publishing},
}

@article{peter_thickness-dependent_2006,
	title = {Thickness-dependent reduction of the glass-transition temperature in thin polymer films with a free surface},
	volume = {44},
	rights = {Copyright © 2006 Wiley Periodicals, Inc.},
	issn = {1099-0488},
	url = {https://onlinelibrary.wiley.com/doi/abs/10.1002/polb.20924},
	doi = {10.1002/polb.20924},
	pages = {2951--2967},
	number = {20},
	journaltitle = {Journal of Polymer Science Part B: Polymer Physics},
	author = {Peter, Simone and Meyer, Hendrik and Baschnagel, Jörg},
	urldate = {2023-12-14},
	date = {2006},
	langid = {english},
	note = {\_eprint: https://onlinelibrary.wiley.com/doi/pdf/10.1002/polb.20924},
}

@article{hao_mobility_2021,
	title = {Mobility gradients yield rubbery surfaces on top of polymer glasses},
	volume = {596},
	rights = {2021 The Author(s), under exclusive licence to Springer Nature Limited},
	issn = {1476-4687},
	url = {https://www.nature.com/articles/s41586-021-03733-7},
	doi = {10.1038/s41586-021-03733-7},
	pages = {372--376},
	number = {7872},
	journaltitle = {Nature},
	author = {Hao, Zhiwei and Ghanekarade, Asieh and Zhu, Ningtao and Randazzo, Katelyn and Kawaguchi, Daisuke and Tanaka, Keiji and Wang, Xinping and Simmons, David S. and Priestley, Rodney D. and Zuo, Biao},
	urldate = {2023-11-06},
	date = {2021-08},
	langid = {english},
	note = {Number: 7872
Publisher: Nature Publishing Group},
}

@article{Christie_Register_Priestley_2018, title={Direct Measurement of the Local Glass Transition in Self-Assembled Copolymers with Nanometer Resolution}, volume={4}, ISSN={2374-7943}, DOI={10.1021/acscentsci.8b00043}, number={4}, journal={ACS Central Science}, author={Christie, Dane and Register, Richard A. and Priestley, Rodney D.}, year={2018}, pages={504–511} }

@article{colmenero_are_2015,
	title = {Are polymers standard glass-forming systems? The role of intramolecular barriers on the glass-transition phenomena of glass-forming polymers},
	volume = {27},
	issn = {0953-8984},
	url = {https://dx.doi.org/10.1088/0953-8984/27/10/103101},
	doi = {10.1088/0953-8984/27/10/103101},
	shorttitle = {Are polymers standard glass-forming systems?},
	pages = {103101},
	number = {10},
	journaltitle = {Journal of Physics: Condensed Matter},
	shortjournal = {J. Phys.: Condens. Matter},
	author = {Colmenero, J.},
	urldate = {2023-09-20},
	date = {2015-01},
	langid = {english},
	note = {Publisher: {IOP} Publishing},
}

@article{cheng_design_2018,
	title = {Design Rules for Highly Conductive Polymeric Ionic Liquids from Molecular Dynamics Simulations},
	volume = {51},
	issn = {0024-9297},
	url = {https://doi.org/10.1021/acs.macromol.8b00572},
	doi = {10.1021/acs.macromol.8b00572},
	pages = {6630--6644},
	number = {17},
	journaltitle = {Macromolecules},
	shortjournal = {Macromolecules},
	author = {Cheng, Yizi and Yang, Junhong and Hung, Jui-Hsiang and Patra, Tarak K. and Simmons, David S.},
	urldate = {2018-08-22},
	date = {2018},
}

@article{hsu_glass-transition_2016,
	title = {Glass-Transition and Side-Chain Dynamics in Thin Films: Explaining Dissimilar Free Surface Effects for Polystyrene vs Poly(methyl methacrylate)},
	volume = {5},
	url = {https://doi.org/10.1021/acsmacrolett.6b00037},
	doi = {10.1021/acsmacrolett.6b00037},
	shorttitle = {Glass-Transition and Side-Chain Dynamics in Thin Films},
	pages = {481--486},
	number = {4},
	journaltitle = {{ACS} Macro Letters},
	shortjournal = {{ACS} Macro Lett.},
	author = {Hsu, David D. and Xia, Wenjie and Song, Jake and Keten, Sinan},
	urldate = {2023-03-17},
	date = {2016-04-19},
	note = {Publisher: American Chemical Society},
}

@article{marvin_nanoconfinement_2014,
	title = {Nanoconfinement effects on the fragility of glass formation of a model freestanding polymer film},
	volume = {10},
	issn = {1744-6848},
	url = {http://pubs.rsc.org/en/content/articlelanding/2014/sm/c3sm53160k},
	doi = {10.1039/C3SM53160K},
	pages = {3166--3170},
	number = {18},
	journaltitle = {Soft Matter},
	shortjournal = {Soft Matter},
	author = {Marvin, M. D. and Lang, R. J. and Simmons, D. S.},
	urldate = {2014-03-26},
	date = {2014-03-26},
	langid = {english},
}

@article{hung_heterogeneous_2018,
	title = {Heterogeneous Rouse Model Predicts Polymer Chain Translational Normal Mode Decoupling},
	volume = {51},
	issn = {0024-9297},
	url = {https://doi.org/10.1021/acs.macromol.8b00135},
	doi = {10.1021/acs.macromol.8b00135},
	pages = {2887--2898},
	number = {8},
	journaltitle = {Macromolecules},
	shortjournal = {Macromolecules},
	author = {Hung, Jui-Hsiang and Mangalara, Jayachandra Hari and Simmons, David S.},
	urldate = {2018-10-27},
	date = {2018-04-24},
}

@article{lang_combined_2014,
	title = {Combined Dependence of Nanoconfined Tg on Interfacial Energy and Softness of Confinement},
	volume = {3},
	url = {http://dx.doi.org/10.1021/mz500361v},
	doi = {10.1021/mz500361v},
	pages = {758--762},
	number = {8},
	journaltitle = {{ACS} Macro Letters},
	shortjournal = {{ACS} Macro Lett.},
	author = {Lang, Ryan J. and Merling, Weston L. and Simmons, David S.},
	urldate = {2014-11-24},
	date = {2014-08-19},
}

@article{hanakata_local_2012,
	title = {Local variation of fragility and glass transition temperature of ultra-thin supported polymer films},
	volume = {137},
	issn = {00219606},
	url = {http://jcp.aip.org/resource/1/jcpsa6/v137/i24/p244901_s1},
	doi = {doi:10.1063/1.4772402},
	pages = {244901},
	number = {24},
	journaltitle = {The Journal of Chemical Physics},
	author = {Hanakata, Paul Z. and Douglas, Jack F. and Starr, Francis W.},
	urldate = {2013-05-29},
	date = {2012-12-27},
}

@article{yang_glassy_2011,
	title = {Glassy dynamics and mechanical response in dense fluids of soft repulsive spheres. {II}. {Shear} modulus, relaxation-elasticity connections, and rheology},
	volume = {134},
	issn = {00219606},
	url = {http://ezproxy.uakron.edu:2048/login?url=http://search.ebscohost.com/login.aspx?direct=true&db=a9h&AN=74465986&site=ehost-live},
	doi = {10.1063/1.3592565},
	number = {20},
	urldate = {2015-05-04},
	journal = {Journal of Chemical Physics},
	author = {Yang, Jian and Schweizer, Kenneth S.},
	month = may,
	year = {2011},
	pages = {204909},
}

@article{van_zanten_brownian_2000,
	title = {Brownian motion in a single relaxation time {Maxwell} fluid},
	volume = {62},
	issn = {1063-651X},
	url = {file://d:/davedrive/nist/papers/NIST papers.data/PDF/vanZantenJH_PhysRevE_2000-3987013376/vanZantenJH_PhysRevE_2000.pdf},
	doi = {Article},
	number = {4},
	journal = {Physical Review E},
	author = {van Zanten, J. H. and Rufener, K. P.},
	year = {2000},
	pages = {5389--5396},
}

@article{cavagna_supercooled_2009,
	title = {Supercooled liquids for pedestrians},
	volume = {476},
	issn = {0370-1573},
	doi = {10.1016/j.physrep.2009.03.003},
	language = {English},
	number = {4-6},
	journal = {Physics Reports-Review Section of Physics Letters},
	author = {Cavagna, Andrea},
	month = jun,
	year = {2009},
	note = {WOS:000267351200001},
	pages = {51--124},
}

@article{debenedetti_supercooled_2001,
	title = {Supercooled liquids and the glass transition},
	volume = {410},
	url = {file://d:/davedrive/nist/papers/NIST papers.data/PDF/DebenedettiPG_Nature_2001-3272816128/DebenedettiPG_Nature_2001.pdf},
	journal = {Nature},
	author = {Debenedetti, P. G. and Stillinger, F. H.},
	year = {2001},
	pages = {259--267},
}

@article{ghanekarade_signature_2023,
	title = {Signature of collective elastic glass physics in surface-induced long-range tails in dynamical gradients},
	copyright = {2023 The Author(s), under exclusive licence to Springer Nature Limited},
	issn = {1745-2481},
	url = {https://www.nature.com/articles/s41567-023-01995-8},
	doi = {10.1038/s41567-023-01995-8},
	language = {en},
	urldate = {2023-03-23},
	journal = {Nature Physics},
	author = {Ghanekarade, Asieh and Phan, Anh D. and Schweizer, Kenneth S. and Simmons, David S.},
	month = mar,
	year = {2023},
	note = {Publisher: Nature Publishing Group},
	pages = {1--7},
}

@article{mangalara_tuning_2015,
	title = {Tuning {Polymer} {Glass} {Formation} {Behavior} and {Mechanical} {Properties} with {Oligomeric} {Diluents} of {Varying} {Stiffness}},
	volume = {4},
	url = {http://dx.doi.org/10.1021/acsmacrolett.5b00635},
	doi = {10.1021/acsmacrolett.5b00635},
	number = {10},
	urldate = {2015-09-23},
	journal = {ACS Macro Letters},
	author = {Mangalara, Jayachandra Hari and Simmons, David S.},
	month = sep,
	year = {2015},
	pages = {1134--1138},
}

@article{xia_molecular_2015,
	title = {Molecular {Weight} {Effects} on the {Glass} {Transition} and {Confinement} {Behavior} of {Polymer} {Thin} {Films}},
	volume = {36},
	issn = {1022-1336},
	doi = {10.1002/marc.201500194},
	language = {English},
	number = {15},
	journal = {Macromolecular Rapid Communications},
	author = {Xia, Wenjie and Hsu, David D. and Keten, Sinan},
	month = aug,
	year = {2015},
	note = {WOS:000359791700006},
	pages = {1422--1427},
}

@book{rosen_fundamental_1993,
	address = {New York},
	edition = {2nd edition},
	title = {Fundamental Principles of Polymeric Materials},
	isbn = {978-0-471-57525-2},
	publisher = {Wiley-Interscience},
	author = {Rosen, Stephen L.},
	year = {1993},
}

@book{mathot_calorimetry_1994,
	address = {Munich ; New York : Cincinnati},
	title = {Calorimetry and Thermal Analysis of Polymers},
	isbn = {978-1-56990-126-7},
	publisher = {Hanser Pub Inc},
	editor = {Mathot, Vincent B. F. and Benoist, L.},
	year = {1994},
}

@book{rudin_elements_2012,
	address = {San Diego},
	edition = {3rd edition},
	title = {The Elements of Polymer Science and Engineering: An Introductory Text and Reference for Engineers and Chemists},
	isbn = {978-0-12-382178-2},
	shorttitle = {The Elements of Polymer Science and Engineering},
	publisher = {Academic Press},
	author = {Rudin, Alfred and P.Eng, Phillip Choi Ph D.},
	month = dec,
	year = {2012},
}

@book{coleman_fundamentals_1998,
	address = {Lancaster, Pa},
	edition = {2nd Edition},
	title = {Fundamentals of {Polymer} {Science}: {An} {Introductory} {Text}},
	isbn = {978-1-56676-559-6},
	shorttitle = {Fundamentals of {Polymer} {Science}},
	publisher = {CRC Press},
	author = {Coleman, Michael M. and Painter, Paul C.},
	year = {1998},
}

@article{buchenau_relation_1992,
	title = {A {RELATION} {BETWEEN} {FAST} {AND} {SLOW} {MOTIONS} {IN} {GLASSY} {AND} {LIQUID} {SELENIUM}},
	volume = {18},
	issn = {0295-5075},
	url = {file://d:/davedrive/nist/papers/NIST papers.data/PDF/BuchenauAP_EurophysLett_1992-1258443783/BuchenauAP_EurophysLett_1992.pdf},
	doi = {Article},
	number = {6},
	journal = {Europhysics Letters},
	author = {Buchenau, U. and Zorn, R.},
	year = {1992},
	pages = {523--528},
}

@article{hall_aperiodic_1987,
	title = {The aperiodic crystal picture and free-energy barriers in glasses},
	volume = {86},
	issn = {0021-9606},
	url = {file://d:/davedrive/nist/papers/NIST papers.data/PDF/HallRW_JChemPhys_1987-2719609104/HallRW_JChemPhys_1987.pdf},
	doi = {Article},
	number = {5},
	journal = {Journal of Chemical Physics},
	author = {Hall, R. W. and Wolynes, P. G.},
	year = {1987},
	pages = {2943--2948},
}

@article{dyre_colloquium:_2006,
	title = {Colloquium: {The} glass transition and elastic models of glass-forming liquids},
	volume = {78},
	issn = {0034-6861},
	url = {file://d:/davedrive/nist/papers/NIST papers.data/PDF/DyreJC_RevModPhys_2006-3254851591/DyreJC_RevModPhys_2006.pdf},
	doi = {Review},
	number = {3},
	journal = {Reviews of Modern Physics},
	author = {Dyre, J. C.},
	year = {2006},
	pages = {953--972},
}

@article{doolittle_studies_1951,
	title = {Studies in {Newtonian} {Flow}. {II}. {The} {Dependence} of the {Viscosity} of {Liquids} on {Free}‐{Space}},
	volume = {22},
	issn = {0021-8979, 1089-7550},
	url = {http://scitation.aip.org/content/aip/journal/jap/22/12/10.1063/1.1699894},
	doi = {10.1063/1.1699894},
	number = {12},
	urldate = {2016-05-09},
	journal = {Journal of Applied Physics},
	author = {Doolittle, Arthur K.},
	month = dec,
	year = {1951},
	pages = {1471--1475},
}

@article{mirigian_elastically_2014-2,
	title = {Elastically cooperative activated barrier hopping theory of relaxation in viscous fluids. {II}. {Thermal} liquids},
	volume = {140},
	issn = {0021-9606},
	url = {https://aip.scitation.org/doi/10.1063/1.4874843},
	doi = {10.1063/1.4874843},
	number = {19},
	urldate = {2021-10-12},
	journal = {The Journal of Chemical Physics},
	author = {Mirigian, Stephen and Schweizer, Kenneth S.},
	month = may,
	year = {2014},
	note = {Publisher: American Institute of Physics},
	pages = {194507},
}

@article{mirigian_elastically_2014,
	title = {Elastically cooperative activated barrier hopping theory of relaxation in viscous fluids. {I}. {General} formulation and application to hard sphere fluids},
	volume = {140},
	issn = {0021-9606},
	doi = {10.1063/1.4874842},
	language = {English},
	number = {19},
	journal = {Journal of Chemical Physics},
	author = {Mirigian, Stephen and Schweizer, Kenneth S.},
	month = may,
	year = {2014},
	note = {WOS:000336832700028},
	pages = {194506},
}

@article{white_explaining_2017,
	title = {Explaining the {T},{V}-dependent dynamics of glass forming liquids: {The} cooperative free volume model tested against new simulation results},
	volume = {147},
	issn = {0021-9606},
	shorttitle = {Explaining the {T},{V}-dependent dynamics of glass forming liquids},
	url = {http://aip.scitation.org/doi/10.1063/1.5001714},
	doi = {10.1063/1.5001714},
	number = {18},
	urldate = {2018-02-02},
	journal = {The Journal of Chemical Physics},
	author = {White, Ronald P. and Lipson, Jane E. G.},
	month = nov,
	year = {2017},
	pages = {184503},
}

@article{adam_temperature_1965,
	title = {On the {Temperature} {Dependence} of {Cooperative} {Relaxation} {Properties} in {Glass}‐{Forming} {Liquids}},
	volume = {43},
	issn = {0021-9606, 1089-7690},
	url = {http://aip.scitation.org/doi/10.1063/1.1696442},
	doi = {10.1063/1.1696442},
	language = {en},
	number = {1},
	urldate = {2020-07-02},
	journal = {The Journal of Chemical Physics},
	author = {Adam, Gerold and Gibbs, Julian H.},
	month = jul,
	year = {1965},
	pages = {139--146},
}

@article{zhou_activated_2022,
	title = {Activated relaxation in supercooled monodisperse atomic and polymeric {WCA} fluids: {Simulation} and {ECNLE} theory},
	volume = {156},
	issn = {0021-9606},
	shorttitle = {Activated relaxation in supercooled monodisperse atomic and polymeric {WCA} fluids},
	url = {https://aip.scitation.org/doi/full/10.1063/5.0079221},
	doi = {10.1063/5.0079221},
	number = {11},
	urldate = {2023-03-15},
	journal = {The Journal of Chemical Physics},
	author = {Zhou, Yuxing and Mei, Baicheng and Schweizer, Kenneth S.},
	month = mar,
	year = {2022},
	note = {Publisher: American Institute of Physics},
	pages = {114901},
}

@article{novikov_temperature_2022,
	title = {Temperature {Dependence} of {Structural} {Relaxation} in {Glass}-{Forming} {Liquids} and {Polymers}},
	volume = {24},
	copyright = {http://creativecommons.org/licenses/by/3.0/},
	issn = {1099-4300},
	url = {https://www.mdpi.com/1099-4300/24/8/1101},
	doi = {10.3390/e24081101},
	language = {en},
	number = {8},
	urldate = {2023-02-08},
	journal = {Entropy},
	author = {Novikov, Vladimir N. and Sokolov, Alexei P.},
	month = aug,
	year = {2022},
	note = {Number: 8
Publisher: Multidisciplinary Digital Publishing Institute},
	pages = {1101},
}

@article{sokolov_why_2007,
	title = {Why many polymers are so fragile},
	volume = {19},
	issn = {0953-8984},
	url = {https://dx.doi.org/10.1088/0953-8984/19/20/205116},
	doi = {10.1088/0953-8984/19/20/205116},
	language = {en},
	number = {20},
	urldate = {2023-03-27},
	journal = {Journal of Physics: Condensed Matter},
	author = {Sokolov, A. P. and Novikov, V. N. and Ding, Y.},
	month = apr,
	year = {2007},
	pages = {205116},
}

@article{cown_general_1975,
	title = {Some general features of {Tg}-{M} relations for oligomers and amorphous polymers},
	volume = {11},
	issn = {0014-3057},
	url = {https://www.sciencedirect.com/science/article/pii/0014305775900373},
	doi = {10.1016/0014-3057(75)90037-3},
	language = {en},
	number = {4},
	urldate = {2023-06-13},
	journal = {European Polymer Journal},
	author = {Cown, J. M. G.},
	month = apr,
	year = {1975},
	pages = {297--300},
}

@article{zaccone_disorder-assisted_2013,
	title = {Disorder-Assisted Melting and the Glass Transition in Amorphous Solids},
	volume = {110},
	url = {https://link.aps.org/doi/10.1103/PhysRevLett.110.178002},
	doi = {10.1103/PhysRevLett.110.178002},
	pages = {178002},
	number = {17},
	journal = {Physical Review Letters},
	shortjournal = {Phys. Rev. Lett.},
	author = {Zaccone, Alessio and Terentjev, Eugene M.},
	urldate = {2023-03-28},
	date = {2013-04-26},
	year = {2013-04-26},
	note = {Publisher: American Physical Society},
}

@article{jorgensen_development_1996,
	title = {Development and Testing of the {OPLS} All-Atom Force Field on Conformational Energetics and Properties of Organic Liquids},
	volume = {118},
	issn = {0002-7863},
	url = {http://dx.doi.org/10.1021/ja9621760},
	doi = {10.1021/ja9621760},
	pages = {11225--11236},
	number = {45},
	journaltitle = {Journal of the American Chemical Society},
	journal = {Journal of the American Chemical Society},
	shortjournal = {J. Am. Chem. Soc.},
	author = {Jorgensen, William L. and Maxwell, David S. and Tirado-Rives, Julian},
	urldate = {2015-12-14},
	date = {1996-01-01},
	year = {1996},
}

@article{puosi_fast_2019,
	title = {Fast Vibrational Modes and Slow Heterogeneous Dynamics in Polymers and Viscous Liquids},
	volume = {20},
	rights = {http://creativecommons.org/licenses/by/3.0/},
	issn = {1422-0067},
	url = {https://www.mdpi.com/1422-0067/20/22/5708},
	doi = {10.3390/ijms20225708},
	pages = {5708},
	number = {22},
	journaltitle = {International Journal of Molecular Sciences},
	journal = {International Journal of Molecular Sciences},
	author = {Puosi, Francesco and Tripodo, Antonio and Leporini, Dino},
	urldate = {2023-02-08},
	date = {2019-01},
	year = {2019},
	langid = {english},
	note = {Number: 22
Publisher: Multidisciplinary Digital Publishing Institute},
}

@article{mckenzie-smith_relating_2022,
	title = {Relating dynamic free volume to cooperative relaxation in a glass-forming polymer composite},
	volume = {157},
	issn = {0021-9606},
	url = {https://aip.scitation.org/doi/full/10.1063/5.0114902},
	doi = {10.1063/5.0114902},
	pages = {131101},
	number = {13},
	journaltitle = {The Journal of Chemical Physics},
	journal = {The Journal of Chemical Physics},
	shortjournal = {J. Chem. Phys.},
	author = {{McKenzie}-Smith, Thomas and Douglas, Jack F. and Starr, Francis W.},
	urldate = {2023-02-08},
	date = {2022-10-07},
	year = {2022},
	note = {Publisher: American Institute of Physics},
}

@article{mckenzie-smith_explaining_2021,
	title = {Explaining the Sensitivity of Polymer Segmental Relaxation to Additive Size Based on the Localization Model},
	volume = {127},
	url = {https://link.aps.org/doi/10.1103/PhysRevLett.127.277802},
	doi = {10.1103/PhysRevLett.127.277802},
	pages = {277802},
	number = {27},
	journaltitle = {Physical Review Letters},
	journal = {Physical Review Letters},
	shortjournal = {Phys. Rev. Lett.},
	author = {{McKenzie}-Smith, Thomas Q. and Douglas, Jack F. and Starr, Francis W.},
	urldate = {2023-02-08},
	date = {2021-12-30},
	year = {2021},
	note = {Publisher: American Physical Society},
}

@article{pazmino_betancourt_quantitative_2015,
	title = {Quantitative relations between cooperative motion, emergent elasticity, and free volume in model glass-forming polymer materials},
	volume = {112},
	url = {https://www.pnas.org/doi/full/10.1073/pnas.1418654112},
	doi = {10.1073/pnas.1418654112},
	pages = {2966--2971},
	number = {10},
	journaltitle = {Proceedings of the National Academy of Sciences},
	journal = {Proceedings of the National Academy of Sciences},
	author = {Pazmiño Betancourt, Beatriz A. and Hanakata, Paul Z. and Starr, Francis W. and Douglas, Jack F.},
	urldate = {2023-02-08},
	date = {2015-03-10},
	year = {2015},
	note = {Publisher: Proceedings of the National Academy of Sciences},
}

@article{ngai_correlation_2001,
	title = {Correlation of Positron Annihilation and Other Dynamic Properties in Small Molecule Glass-Forming Substances},
	volume = {87},
	url = {https://link.aps.org/doi/10.1103/PhysRevLett.87.215901},
	doi = {10.1103/PhysRevLett.87.215901},
	pages = {215901},
	number = {21},
	journaltitle = {Physical Review Letters},
	journal = {Physical Review Letters},
	shortjournal = {Phys. Rev. Lett.},
	author = {Ngai, Kia L. and Bao, Li-Rong and Yee, Albert F. and Soles, Christopher L.},
	urldate = {2023-02-08},
	date = {2001-11-05},
	year = {2001},
	note = {Publisher: American Physical Society},
}

@article{ngai_why_2004,
	title = {Why the fast relaxation in the picosecond to nanosecond time range can sense the glass transition},
	volume = {84},
	issn = {1478-6435},
	url = {https://doi.org/10.1080/14786430310001644080},
	doi = {10.1080/14786430310001644080},
	pages = {1341--1353},
	number = {13},
	journaltitle = {Philosophical Magazine},
	journal = {Philosophical Magazine},
	author = {Ngai, K. L.},
	urldate = {2023-02-08},
	date = {2004-05-01},
	year = {2004},
	note = {Publisher: Taylor \& Francis
\_eprint: https://doi.org/10.1080/14786430310001644080},
}

@article{ottochian_universal_2011,
	title = {Universal scaling between structural relaxation and caged dynamics in glass-forming systems: Free volume and time scales},
	volume = {357},
	issn = {0022-3093},
	url = {https://www.sciencedirect.com/science/article/pii/S0022309310005144},
	doi = {10.1016/j.jnoncrysol.2010.05.094},
	series = {6th International Discussion Meeting on Relaxation in Complex Systems},
	shorttitle = {Universal scaling between structural relaxation and caged dynamics in glass-forming systems},
	pages = {298--301},
	number = {2},
	journaltitle = {Journal of Non-Crystalline Solids},
	shortjournal = {Journal of Non-Crystalline Solids},
	journal = {Journal of Non-Crystalline Solids},
	author = {Ottochian, A. and Leporini, D.},
	urldate = {2023-02-06},
	date = {2011-01-15},
	year = {2011},
	langid = {english},
}

@article{miwa_influence_2003,
	title = {Influence of Chain End and Molecular Weight on Molecular Motion of Polystyrene, Revealed by the {ESR} Selective Spin-Label Method},
	volume = {36},
	issn = {0024-9297},
	url = {https://doi.org/10.1021/ma030026l},
	doi = {10.1021/ma030026l},
	pages = {3235--3239},
	number = {9},
	journaltitle = {Macromolecules},
	shortjournal = {Macromolecules},
	journal = {Macromolecules},
	author = {Miwa, Yohei and Tanase, Takayuki and Yamamoto, Katsuhiro and Sakaguchi, Masato and Sakai, Masahiro and Shimada, Shigetaka},
	urldate = {2022-12-13},
	date = {2003-05-01},
	year = {2003},
	note = {Publisher: American Chemical Society},
}

@article{starr_what_2002,
	title = {What Do We Learn from the Local Geometry of Glass-Forming Liquids?},
	volume = {89},
	url = {http://link.aps.org/doi/10.1103/PhysRevLett.89.125501},
	doi = {10.1103/PhysRevLett.89.125501},
	pages = {125501},
	number = {12},
	journal = {Physical Review Letters},
	shortjournal = {Phys. Rev. Lett.},
	author = {Starr, Francis W. and Sastry, Srikanth and Douglas, Jack F. and Glotzer, Sharon C.},
	urldate = {2013-08-23},
	date = {2002-08-28},
	year = {2002},
}

@book{hiemenz_polymer_2007,
	year = {2007},
	edition = {2nd ed},
	title = {Polymer Chemistry},
	isbn = {978-1-57444-779-8},
	publisher = {{CRC} Press},
	author = {Hiemenz, Paul C. and Lodge, Timothy},
	date = {2007},
}

@book{rubinstein_polymer_2003,
	title = {Polymer Physics},
	isbn = {978-0-19-852059-7},
	pagetotal = {458},
	publisher = {{OUP} Oxford},
	author = {Rubinstein, Michael and Colby, Ralph H.},
	date = {2003-06-26},
	year = {2003},
	langid = {english},
}

@article{hung_universal_2019,
	title = {Universal localization transition accompanying glass formation: insights from efficient molecular dynamics simulations of diverse supercooled liquids},
	volume = {15},
	issn = {1744-6848},
	url = {http://pubs.rsc.org/en/content/articlelanding/2019/sm/c8sm02051e},
	doi = {10.1039/C8SM02051E},
	shorttitle = {Universal localization transition accompanying glass formation},
	pages = {1223--1242},
	number = {6},
	journal = {Soft Matter},
	shortjournal = {Soft Matter},
	author = {Hung, Jui-Hsiang and Patra, Tarak K. and Meenakshisundaram, Venkatesh and Mangalara, Jayachandra Hari and Simmons, David S.},
	urldate = {2019-02-24},
	date = {2019-02-06},
	year = {2019},
	langid = {english},
}

@article{thompson_lammps_2022,
	title = {{LAMMPS} - a flexible simulation tool for particle-based materials modeling at the atomic, meso, and continuum scales},
	volume = {271},
	issn = {0010-4655},
	url = {https://www.sciencedirect.com/science/article/pii/S0010465521002836},
	doi = {10.1016/j.cpc.2021.108171},
	pages = {108171},
	journal = {Computer Physics Communications},
	shortjournal = {Computer Physics Communications},
	author = {Thompson, Aidan P. and Aktulga, H. Metin and Berger, Richard and Bolintineanu, Dan S. and Brown, W. Michael and Crozier, Paul S. and in 't Veld, Pieter J. and Kohlmeyer, Axel and Moore, Stan G. and Nguyen, Trung Dac and Shan, Ray and Stevens, Mark J. and Tranchida, Julien and Trott, Christian and Plimpton, Steven J.},
	urldate = {2021-11-12},
	date = {2022-02-01},
	year = {2022},
	langid = {english},
}

@article{mirigian_dynamical_2015,
	title = {Dynamical Theory of Segmental Relaxation and Emergent Elasticity in Supercooled Polymer Melts},
	volume = {48},
	issn = {0024-9297},
	url = {http://dx.doi.org/10.1021/ma5022083},
	doi = {10.1021/ma5022083},
	pages = {1901--1913},
	number = {6},
	journal = {Macromolecules},
	shortjournal = {Macromolecules},
	author = {Mirigian, Stephen and Schweizer, Kenneth S.},
	urldate = {2016-04-28},
	date = {2015-03-24},
	year = {2015},
}

@article{novikov_correlation_2013,
	title = {Correlation between glass transition temperature and molecular mass in non-polymeric and polymer glass formers},
	volume = {54},
	issn = {00323861},
	url = {https://linkinghub.elsevier.com/retrieve/pii/S0032386113010288},
	doi = {10.1016/j.polymer.2013.11.002},
	pages = {6987--6991},
	number = {26},
	journal = {Polymer},
	shortjournal = {Polymer},
	author = {Novikov, V.N. and Rössler, E.A.},
	urldate = {2022-03-02},
	date = {2013-12},
	year = {2013},
	langid = {english},
}

@article{ueberreiter_self-plasticization_1952,
	title = {Self-plasticization of polymers},
	volume = {7},
	issn = {0095-8522},
	url = {https://www.sciencedirect.com/science/article/pii/0095852252900408},
	doi = {10.1016/0095-8522(52)90040-8},
	pages = {569--583},
	number = {6},
	journal = {Journal of Colloid Science},
	shortjournal = {Journal of Colloid Science},
	author = {Ueberreiter, Kurt and Kanig, Gerhard},
	urldate = {2022-12-12},
	date = {1952-12-01},
	year = {1952},
	langid = {english},
}

@article{fox_secondorder_1950,
	title = {Second‐Order Transition Temperatures and Related Properties of Polystyrene. I. Influence of Molecular Weight},
	volume = {21},
	issn = {0021-8979},
	url = {https://aip.scitation.org/doi/10.1063/1.1699711},
	doi = {10.1063/1.1699711},
	pages = {581--591},
	number = {6},
	journal = {Journal of Applied Physics},
	author = {Fox, Thomas G. and Flory, Paul J.},
	urldate = {2022-12-12},
	date = {1950-06},
	year = {1950},
	note = {Publisher: American Institute of Physics},
}

@article{kremer_dynamics_1990,
        title = {Dynamics of entangled linear polymer melts: A molecular-dynamics simulation},
	volume = {92},
	issn = {0021-9606},
	url = {https://aip.scitation.org/doi/10.1063/1.458541},
	doi = {10.1063/1.458541},
	shorttitle = {Dynamics of entangled linear polymer melts},
	pages = {5057--5086},
	number = {8},
	journal = {The Journal of Chemical Physics},
	shortjournal = {J. Chem. Phys.},
	author = {Kremer, Kurt and Grest, Gary S.},
	urldate = {2020-01-22},
	date = {1990-04-15},
	year = {1990},
}

@article{bulacu_molecular-dynamics_2007,
	title = {Molecular-dynamics simulation study of the glass transition in amorphous polymers with controlled chain stiffness},
	volume = {76},
	issn = {1539-3755, 1550-2376},
	url = {https://link.aps.org/doi/10.1103/PhysRevE.76.011807},
	doi = {10.1103/PhysRevE.76.011807},
	pages = {011807},
	number = {1},
	journaltitle = {Physical Review E},
	journal = {Physical Review E},
	shortjournal = {Phys. Rev. E},
	author = {Bulacu, Monica and van der Giessen, Erik},
	urldate = {2022-10-20},
	date = {2007-07-26},
	year = {2007},
	langid = {english},
}

@article{baker_cooperative_2022,
	title = {Cooperative Intra\-molecular Dynamics Control the Chain-Length-Dependent Glass Transition in Polymers},
	volume = {12},
	issn = {2160-3308},
	url = {https://link.aps.org/doi/10.1103/PhysRevX.12.021047},
	doi = {10.1103/PhysRevX.12.021047},
	pages = {021047},
	number = {2},
	journal = {Physical Review X},
	shortjournal = {Phys. Rev. X},
	author = {Baker, Daniel L. and Reynolds, Matthew and Masurel, Robin and Olmsted, Peter D. and Mattsson, Johan},
	urldate = {2022-07-07},
	date = {2022-05-27},
	year = {2022},
	langid = {english},
}

@article{mauro_viscosity_2009,
	title = {Viscosity of glass-forming liquids},
	volume = {106},
	issn = {0027-8424, 1091-6490},
	url = {http://www.pnas.org/content/106/47/19780},
	doi = {10.1073/pnas.0911705106},
	pages = {19780--19784},
	number = {47},
	journal = {Proceedings of the National Academy of Sciences},
	shortjournal = {{PNAS}},
	author = {Mauro, John C. and Yue, Yuanzheng and Ellison, Adam J. and Gupta, Prabhat K. and Allan, Douglas C.},
	urldate = {2014-03-18},
	year = {2009},
	langid = {english},
	pmid = {19903878},
}

@article{schmidtke_temperature_2015,
	title = {Temperature Dependence of the Segmental Relaxation Time of Polymers Revisited},
	volume = {48},
	issn = {0024-9297, 1520-5835},
	url = {https://pubs.acs.org/doi/10.1021/acs.macromol.5b00204},
	doi = {10.1021/acs.macromol.5b00204},
	pages = {3005--3013},
	number = {9},
	journal = {Macromolecules},
	shortjournal = {Macromolecules},
	author = {Schmidtke, B. and Hofmann, M. and Lichtinger, A. and Rössler, E. A.},
	urldate = {2022-03-08},
	date = {2015-05-12},
	year = {2015},
	langid = {english},
}

@article{hung_forecasting_2020,
	title = {Forecasting the experimental glass transition from short time relaxation data},
	year = {2020},
	journal = {Journal of Non-Crystalline Solids},
	volume = {544},
	issn = {0022-3093},
	url = {https://www.sciencedirect.com/science/article/pii/S0022309320303185},
	doi = {10.1016/j.jnoncrysol.2020.120205},
	pages = {120205},
	journaltitle = {Journal of Non-Crystalline Solids},
	shortjournal = {Journal of Non-Crystalline Solids},
	author = {Hung, Jui-Hsiang and Patra, Tarak K. and Simmons, David S.},
	urldate = {2021-09-15},
	date = {2020-09-15},
	langid = {english},
}

\end{document}


\maketitle

\section*{Quantifying Relaxation and its Dependence on Temperature}

\subsection*{Defining Relaxation Times}
\label{choose_tg}

We report here the time evolution of the self-part of the intermediate scattering function, $F_s$, for a chain length of 
$N=10$ and each model studied:
the freely-jointed chain (FJC),
the freely-rotating chain (FRC),
and all-atom polystyrene (AAPS).
We also report for AAPS when $N=100$.
Temperatures are indicated in the legend for each; 
the red data indicate the highest temperature simulated, 
and the blue data indicate our coldest equilibrated temperature.
The data here is only a subset of all temperatures simulated and equilibrated for clarity\footnote{Simulation protocol elaborated further in Hung et al., Soft Matter, 2019, 15, 1223}.
Filled symbols are the mean system;
open symbols are that of the chain ends.

\begin{figure}[htbp]
    \centering
    \includegraphics[height=0.3\textheight]{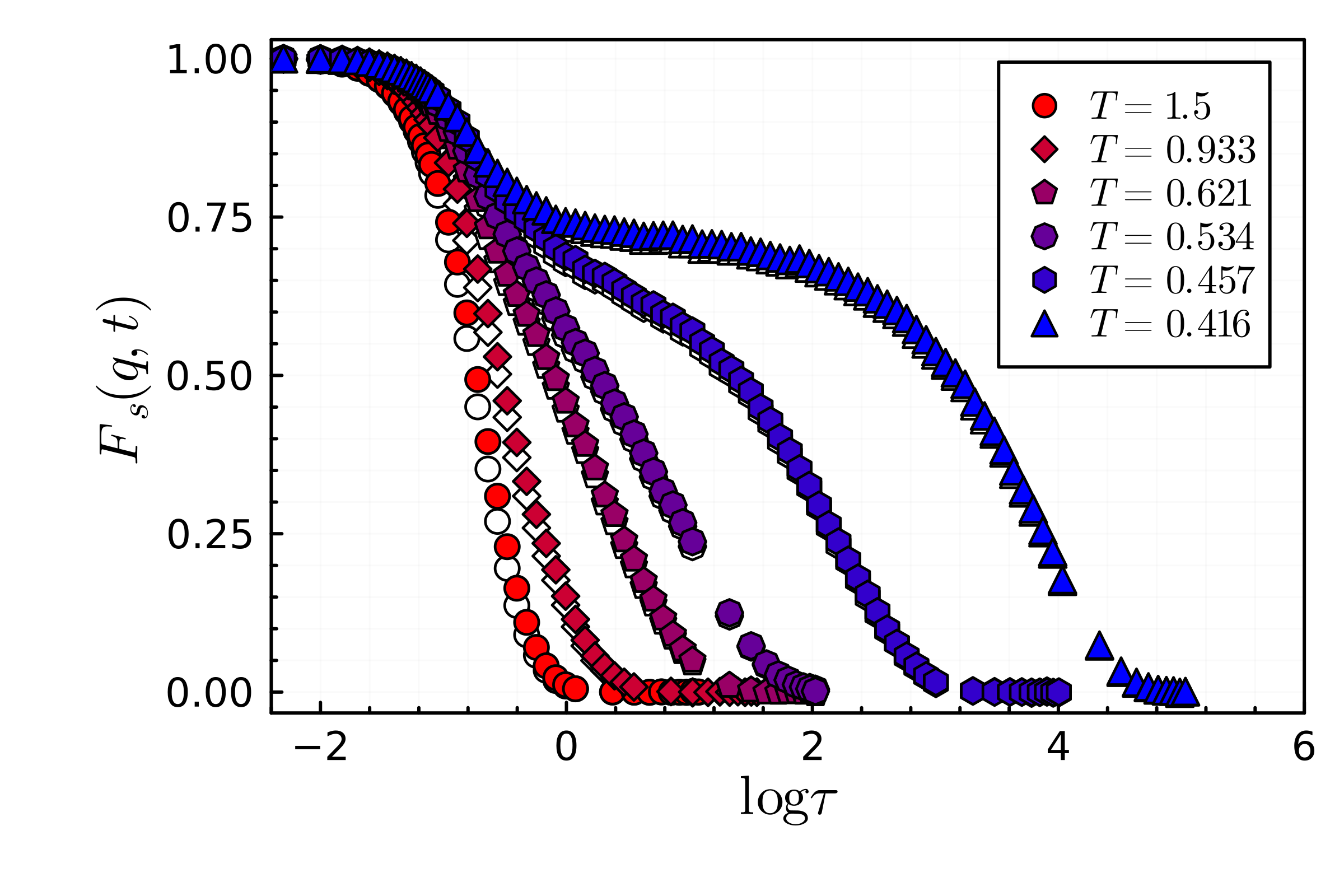}
    \caption{Self-part of the intermediate scattering function for select temperatures
    (roughly a sixth of all temperatures simulated) for the freely-jointed chain de\-ca\-mer ($N=10$);
    the hottest (red circles) and coldest (blue triangles) temperatures simulated are included.
    Chain ends are denoted as open symbols with the corresponding symbol shape as the mean system data.
    }
    \label{fig:isfs_fjc_10}
\end{figure}

\begin{figure}[htbp]
    \centering
    \includegraphics[height=0.3\textheight]{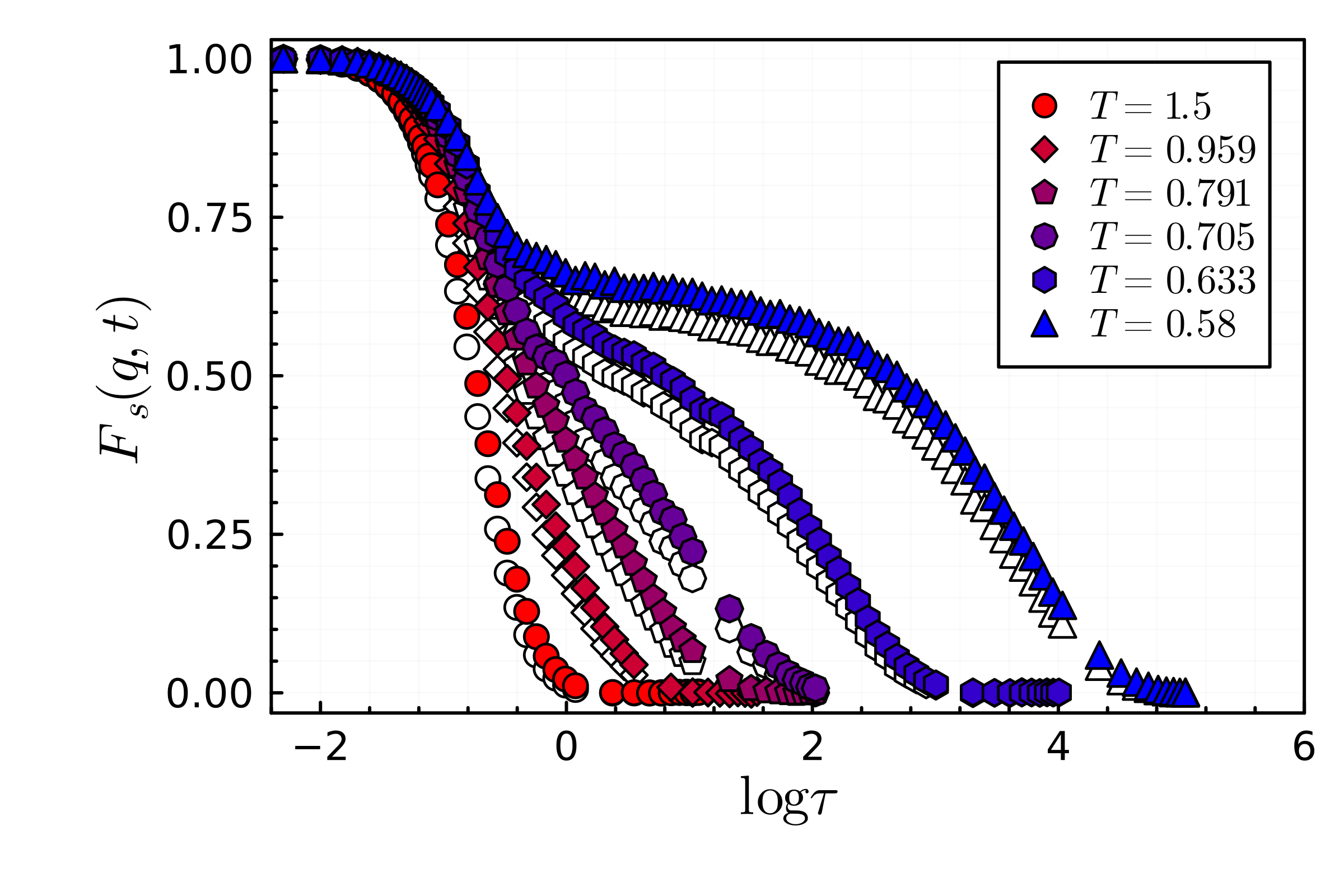}
    \caption{Self-part of the intermediate scattering function for select temperatures
    (roughly a sixth of all temperatures simulated) of the freely-rotating chain de\-ca\-mer ($N=10$);
    the hottest (red circles) and coldest (blue triangles) temperatures simulated are included.
    Chain ends are denoted as open symbols with the corresponding symbol shape as the mean system data.
    }
    \label{fig:isfs_frc_10}
\end{figure}

\begin{figure}[htbp]
    \centering
    \includegraphics[height=0.3\textheight]{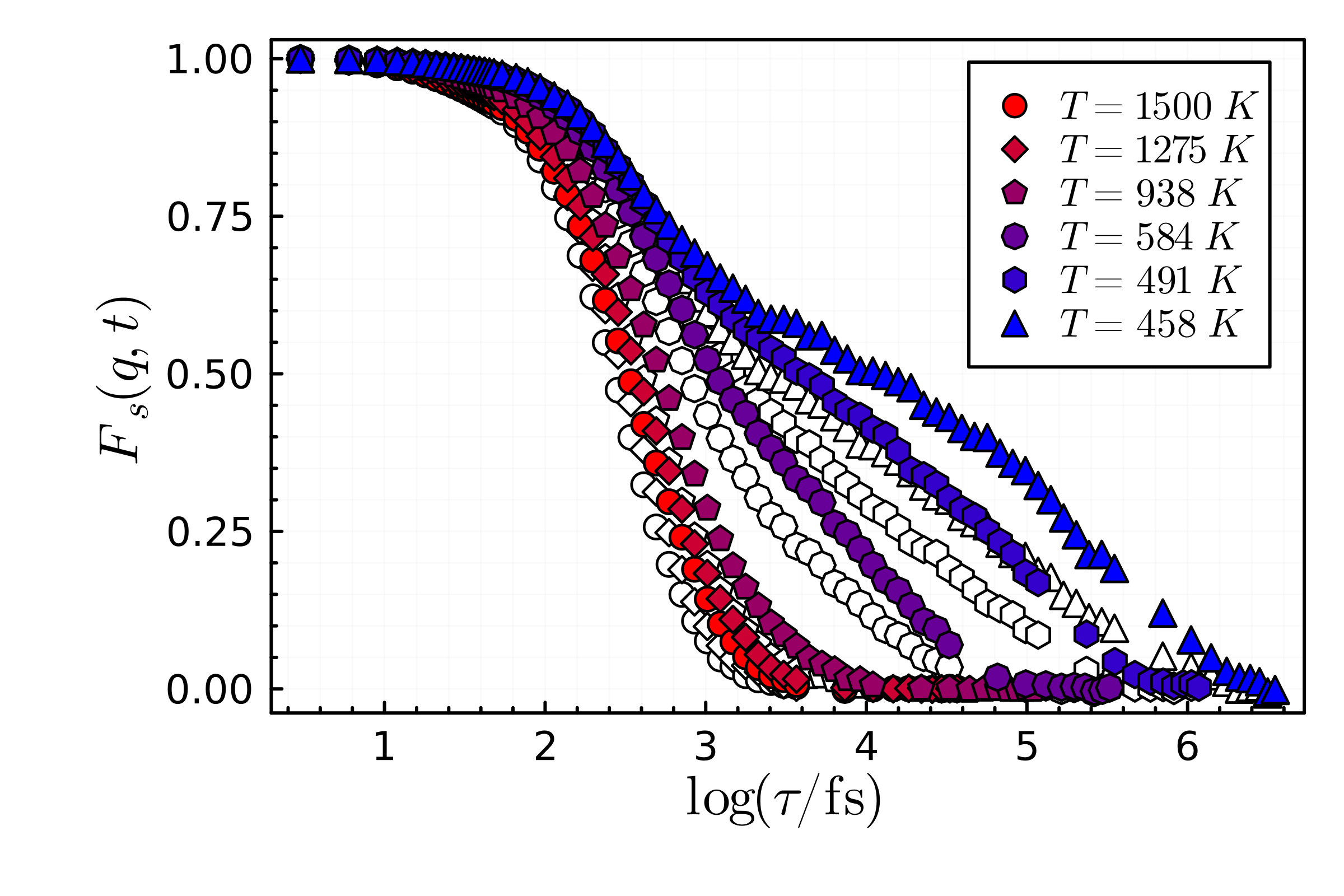}
    \caption{Self-part of the intermediate scattering function for select temperatures
    (roughly a tenth of all temperatures simulated) of the all-atom polystyrene de\-ca\-mer ($N=10$);
    the hottest (red circles) and coldest (blue triangles) temperatures simulated are included.
    Chain ends are denoted as open symbols with the corresponding symbol shape as the mean system data.
    The time unit (before taking the base 10 logarithm) is in femtoseconds.
    }
    \label{fig:isfs_ps_10}
\end{figure}

\begin{figure}[htbp]
    \centering
    \includegraphics[height=0.3\textheight]{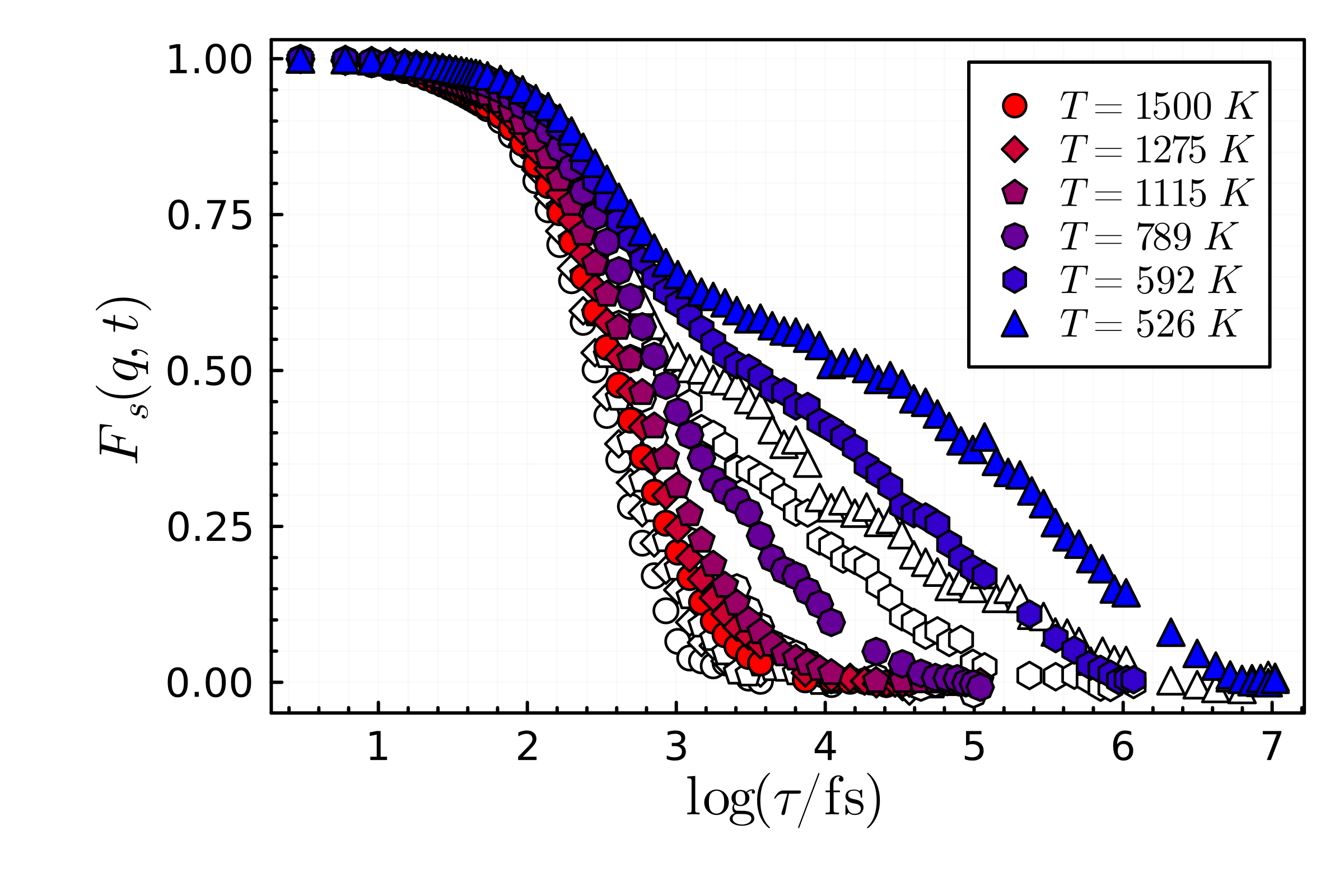}
    \caption{Self-part of the intermediate scattering function for select temperatures
    (roughly a tenth of all temperatures simulated) of the all-atom polystyrene he\-ca\-to\-mer ($N=100$);
    the hottest (red circles) and coldest (blue triangles) temperatures simulated are included.
    Chain ends are denoted as open symbols with the corresponding symbol shape as the mean system data.
    The time unit (before taking the base 10 logarithm) is in femtoseconds.
    }
    \label{fig:isfs_ps_100}
\end{figure}

\clearpage


\subsection*{System Mean and Chain End Relaxation Data}

Here we report figures that demonstrate the chain-end effect 
(or lack thereof)
for relaxation times as a function of inverse temperature for a chain length of $N=10$ for each model 
(along with $N=100$ for AAPS).

\begin{figure}[htbp]
    \centering
    \includegraphics[height=0.3\textheight]{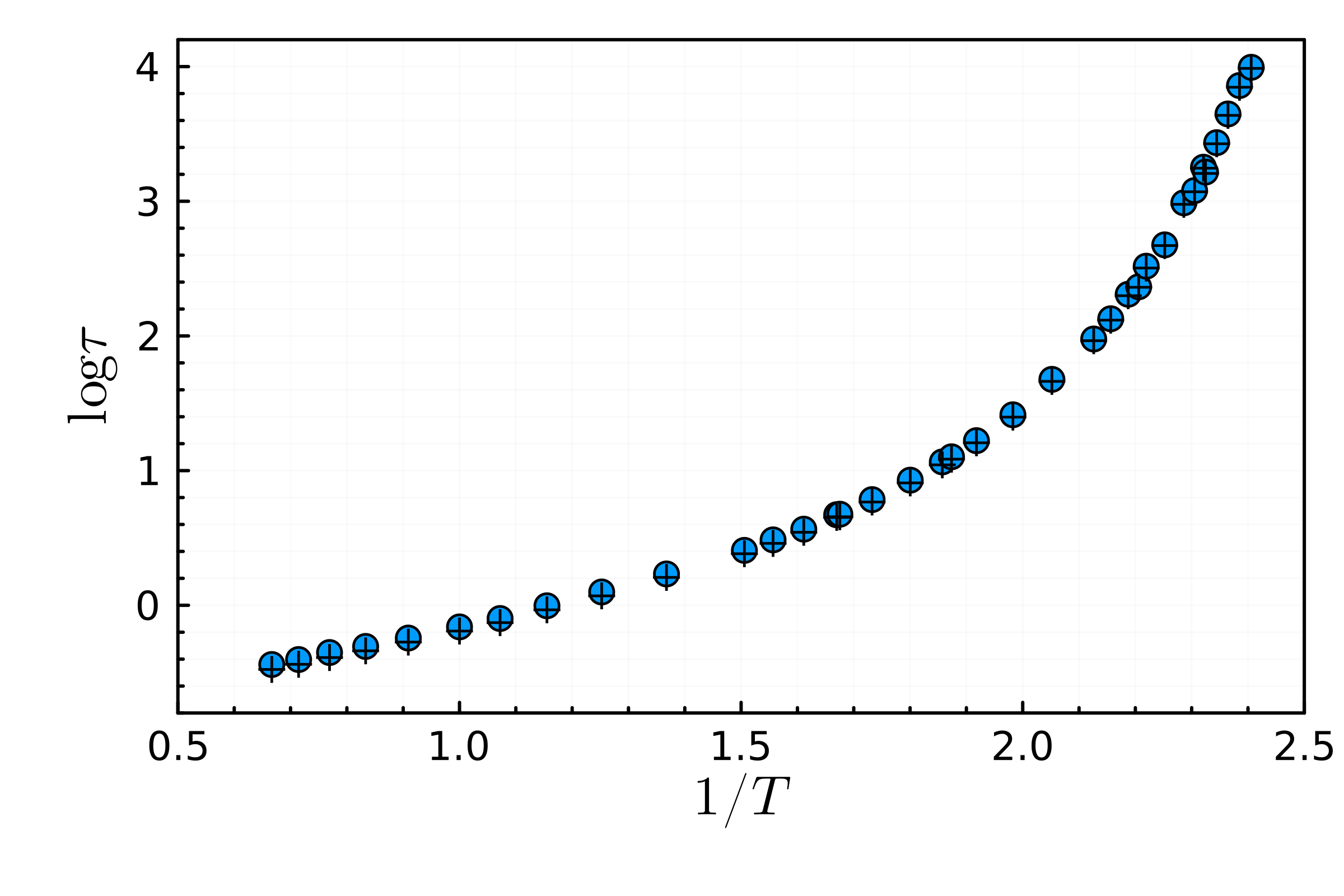}
    \caption{Relaxation time (base 10 logarithm) as a function of inverse temperature for the freely-jointed de\-ca\-mer ($N=10$).
    The mean relaxation time is shown in blue circles, 
    and the chain end relaxation times are in crosses.}
    \label{fig:fjc-10-ends}
\end{figure}

\begin{figure}[htbp]
    \centering
    \includegraphics[height=0.3\textheight]{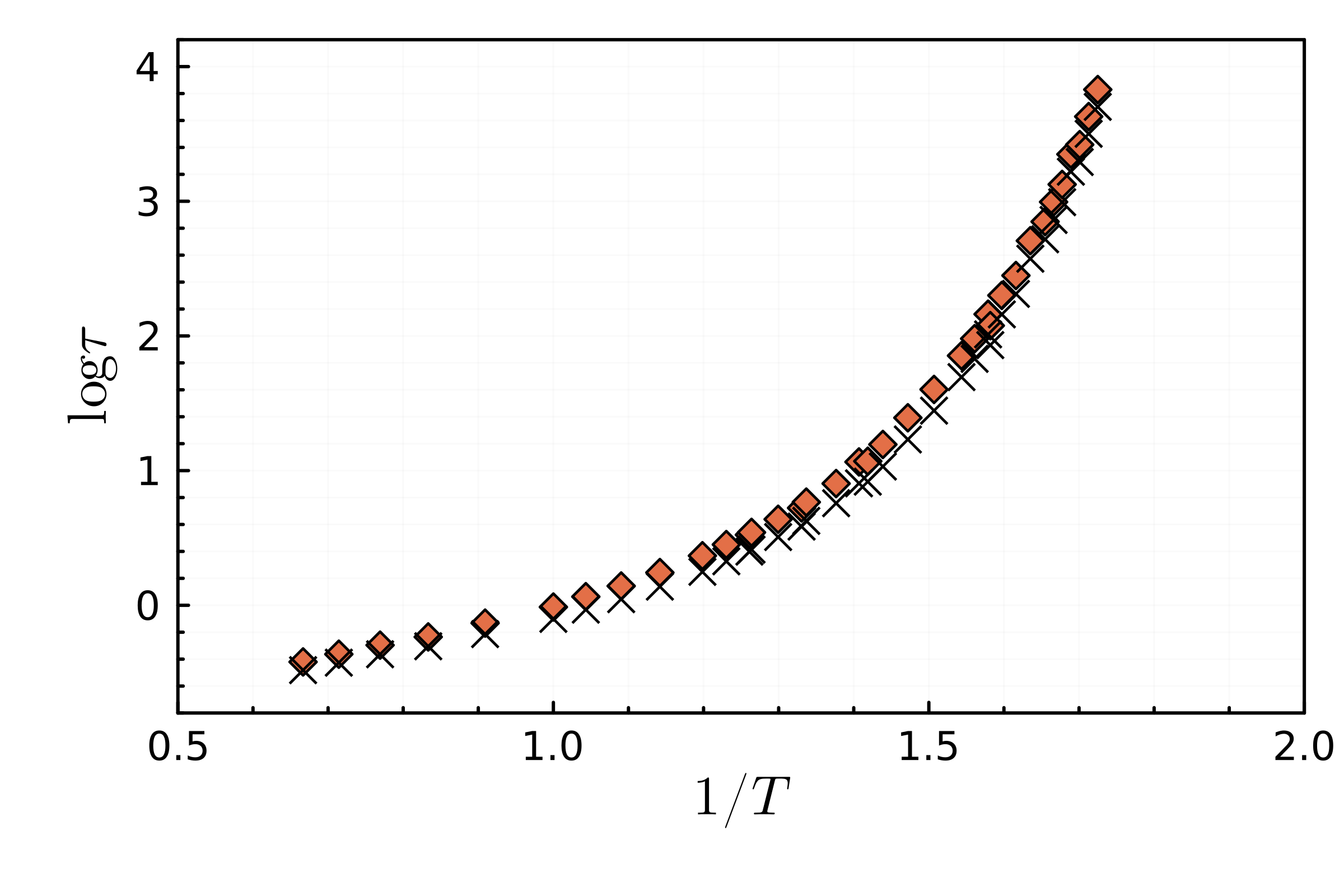}
    \caption{Relaxation time (base 10 logarithm) as a function of inverse temperature for the freely-rotating de\-ca\-mer ($N=10$).
    The mean relaxation time is shown in orange diamonds, 
    and the chain end relaxation times are in crosses.}
    \label{fig:frc-10-ends}
\end{figure}

\begin{figure}[htbp]
    \centering
    \includegraphics[height=0.3\textheight]{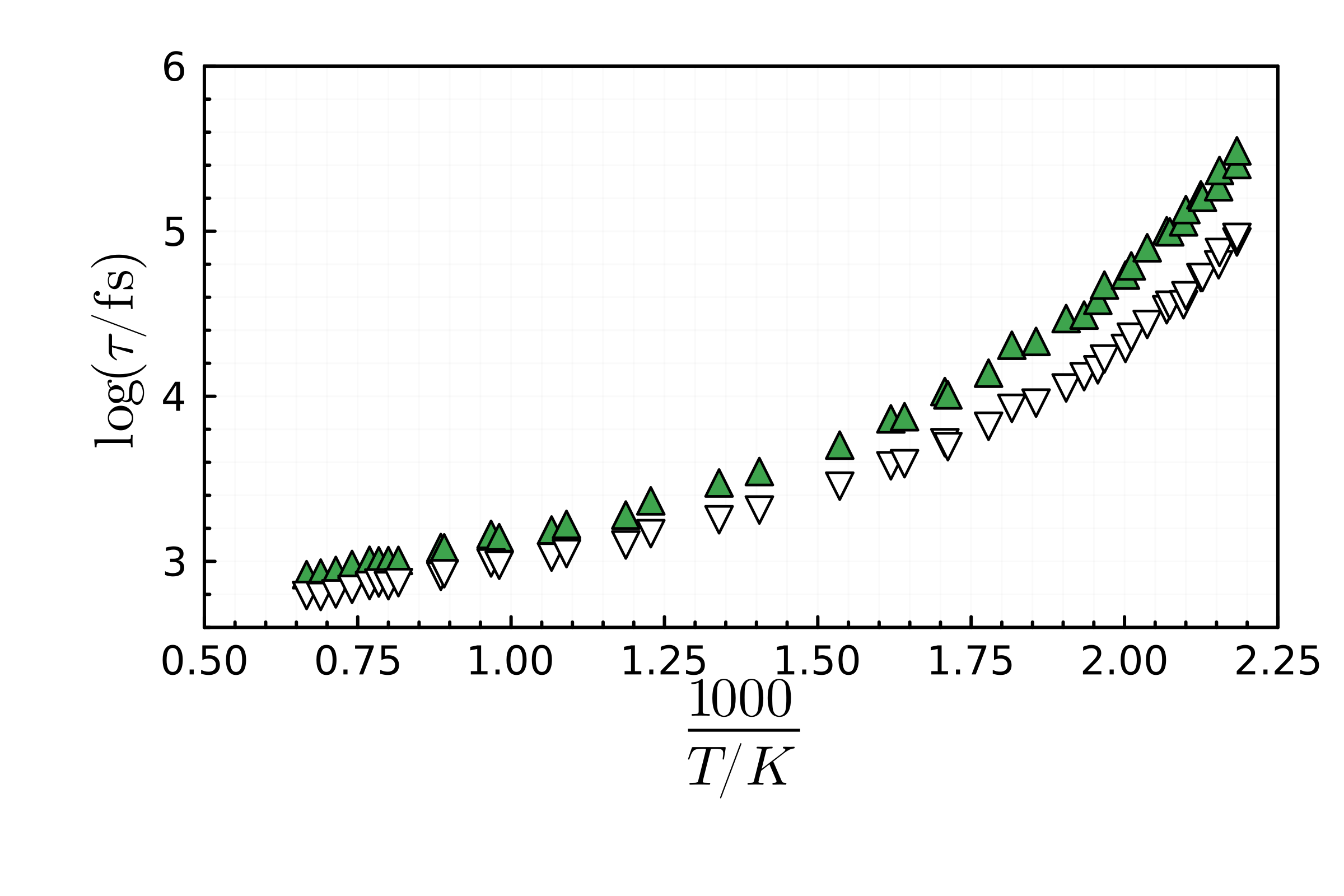}
    \caption{Relaxation time (base 10 logarithm of time in fem\-to\-seconds) as a function of inverse temperature for the all-atom polystyrene de\-ca\-mer ($N=10$).
    The mean relaxation time is shown in green triangles, 
    and the chain end relaxation times are in open triangles.}
    \label{fig:ps-10-ends}
\end{figure}

\begin{figure}[htbp]
    \centering
    \includegraphics[height=0.3\textheight]{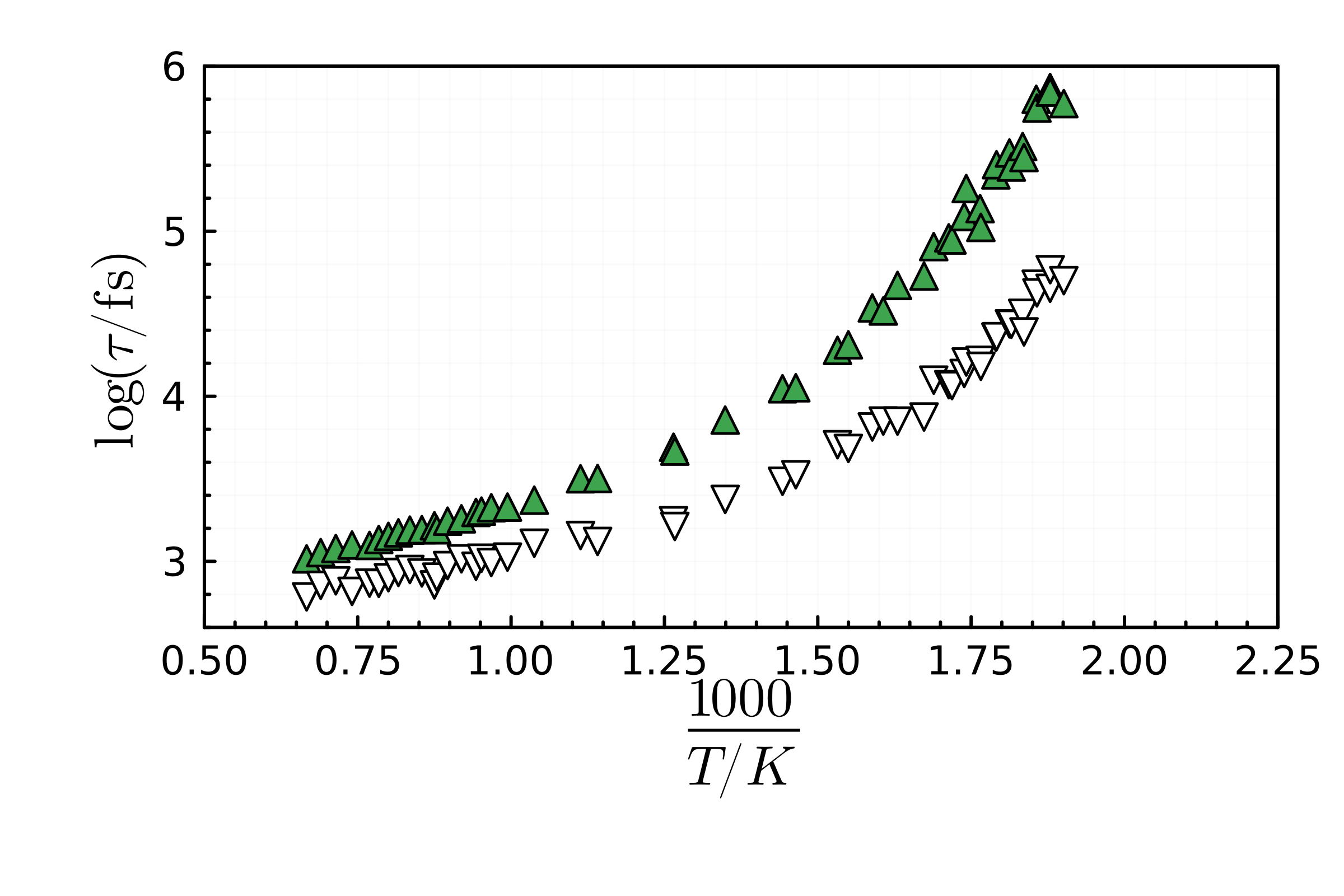}
    \caption{Relaxation time (base 10 logarithm of time in fem\-to\-seconds) as a function of inverse temperature for the all-atom polystyrene he\-ca\-to\-mer ($N=100$).
    The mean relaxation time is shown in green triangles, 
    and the chain end relaxation times are in open triangles.}
    \label{fig:ps-100-ends}
\end{figure}

\newpage

\subsection*{Mean System Relaxation Data Dependence on $N$}

Here we report mean system relaxation times as a function of inverse temperature for each model and all chain lengths studied.

\begin{figure}[htbp]
    \centering
    \includegraphics[height=0.3\textheight]{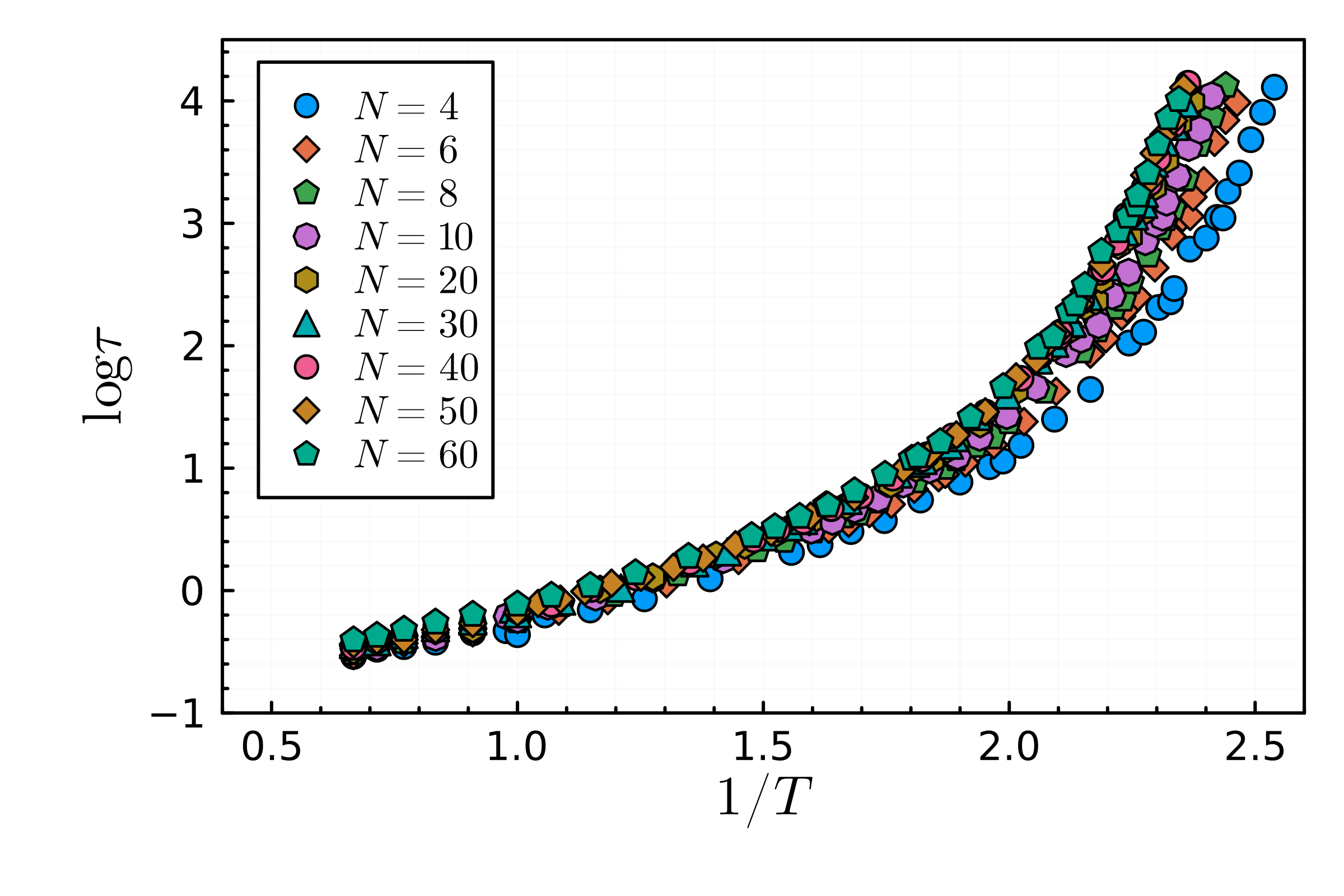}
    \caption{Relaxation time (base 10 logarithm of time in Lennard-Jones  units)  as a function of inverse temperature for all studied chain lengths of the freely-jointed chain.}
    \label{fig:relax_fjc}
\end{figure}

\begin{figure}[htbp]
    \centering
    \includegraphics[height=0.3\textheight]{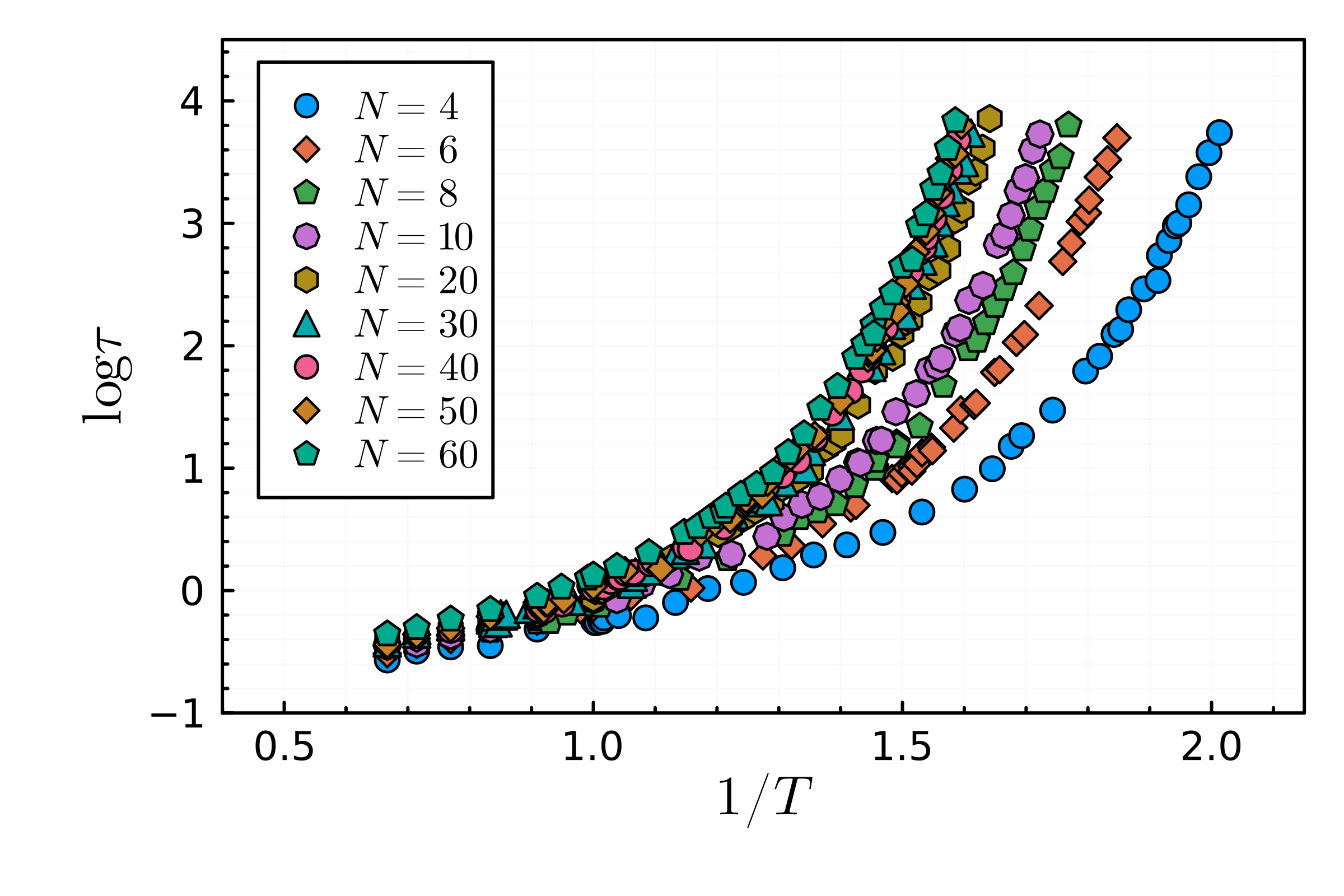}
    \caption{Relaxation time (base 10 logarithm of time in Lennard-Jones units)  as a function of inverse temperature for all studied chain lengths of the freely-rotating chain.}
    \label{fig:relax_frc}
\end{figure}

\begin{figure}[htbp]
    \centering
    \includegraphics[height=0.3\textheight]{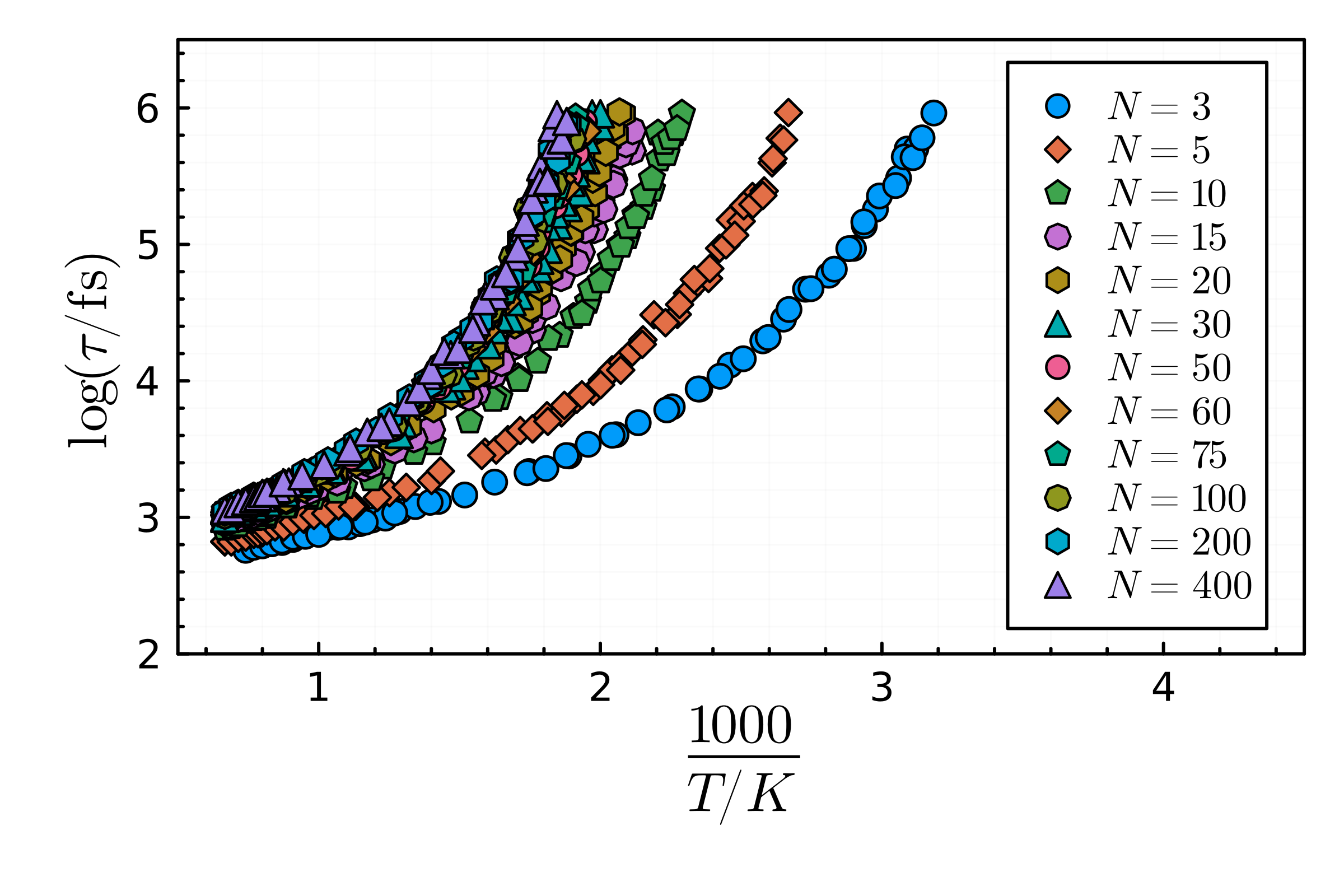}
    \caption{Relaxation time (base 10 logarithm of time in fem\-to\-seconds) as a function of inverse temperature for all studied chain lengths of the all-atom polystyrene (cf. Figure 9b from \cite{hung_forecasting_2020})}
    \label{fig:relax_ps}
\end{figure}

\newpage

\section*{The Debye-Waller Factor and its Dependence on $T$}
\subsection*{Definition of Debye-Waller Factor}

\begin{figure}[htbp]
    \centering
    \includegraphics[height=0.3\textheight]{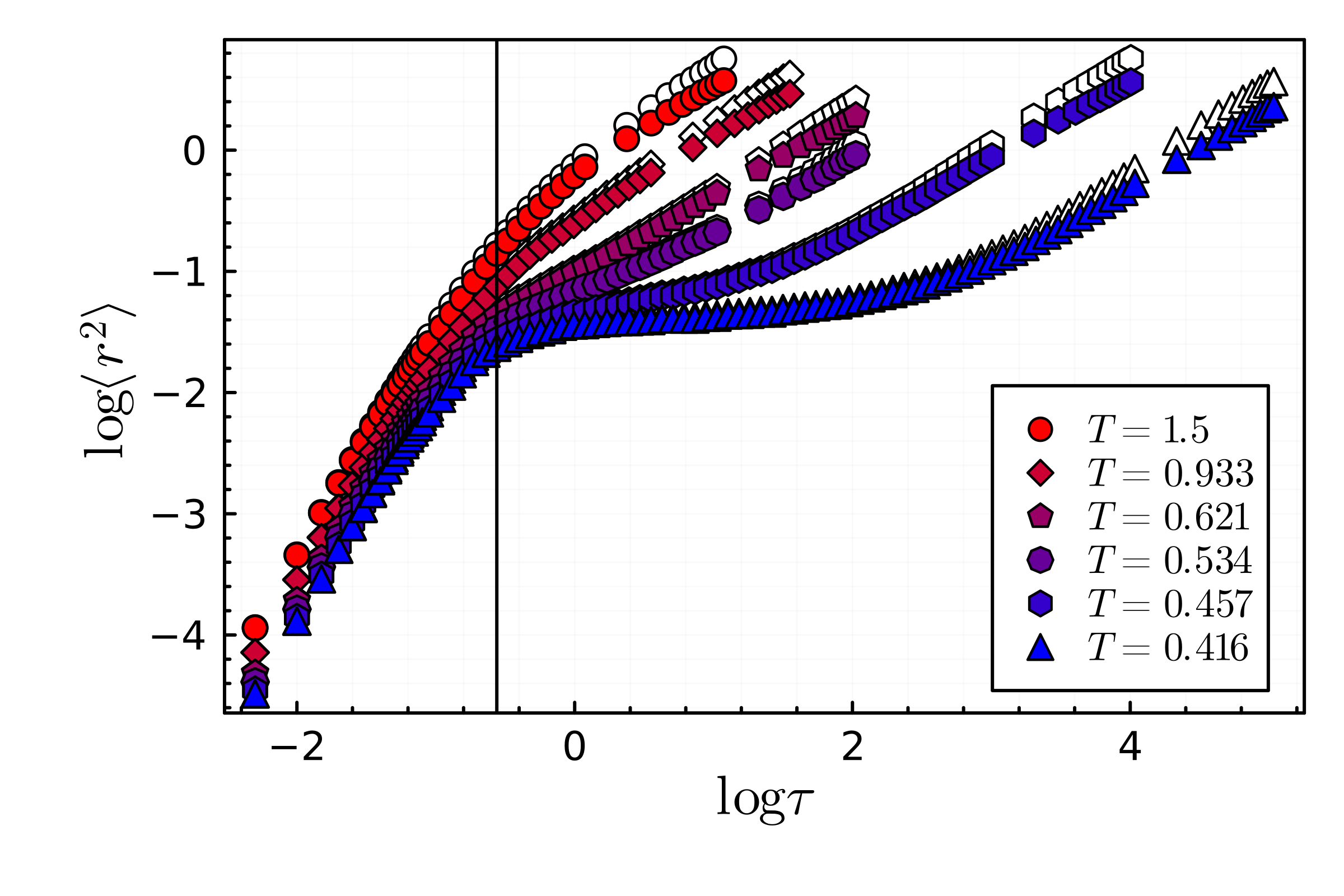}
    \caption{Mean-square displacement data for select temperatures
    (roughly a sixth of all temperatures simulated) of the freely-jointed chain de\-ca\-mer ($N=10$);
    the hottest (red circles) and coldest (blue triangles) temperatures simulated are included.
    Chain ends are denoted as open symbols;
    chain middles are denoted as filled symbols.
    The vertical black line shows the timescale chosen for $\langle u^2 \rangle$.
    }
    \label{fig:msd_fjc_10}
\end{figure}

\begin{figure}[htbp]
    \centering
    \includegraphics[height=0.3\textheight]{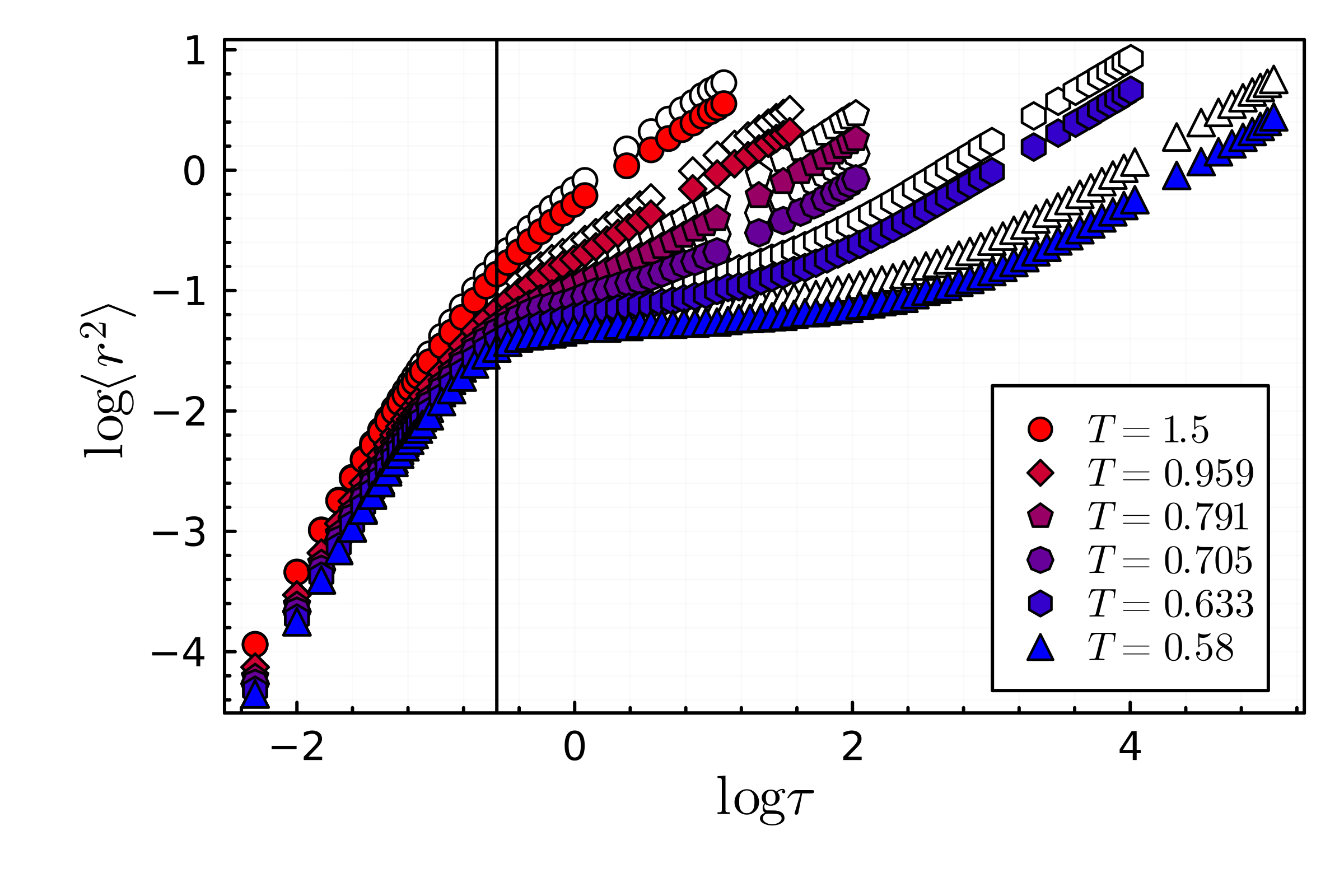}
    \caption{Mean-square displacement data for select temperatures
    (roughly a sixth of all temperatures simulated) of the freely-rotating chain de\-ca\-mer ($N=10$);
    the hottest (red circles) and coldest (blue triangles) temperatures simulated are included.
    Chain ends are denoted as open symbols;
    chain middles are denoted as filled symbols.
    The vertical black line shows the timescale chosen for $\langle u^2 \rangle$.
    }
    \label{fig:msd_frc_10}
\end{figure}

\begin{figure}[htbp]
    \centering
    \includegraphics[height=0.3\textheight]{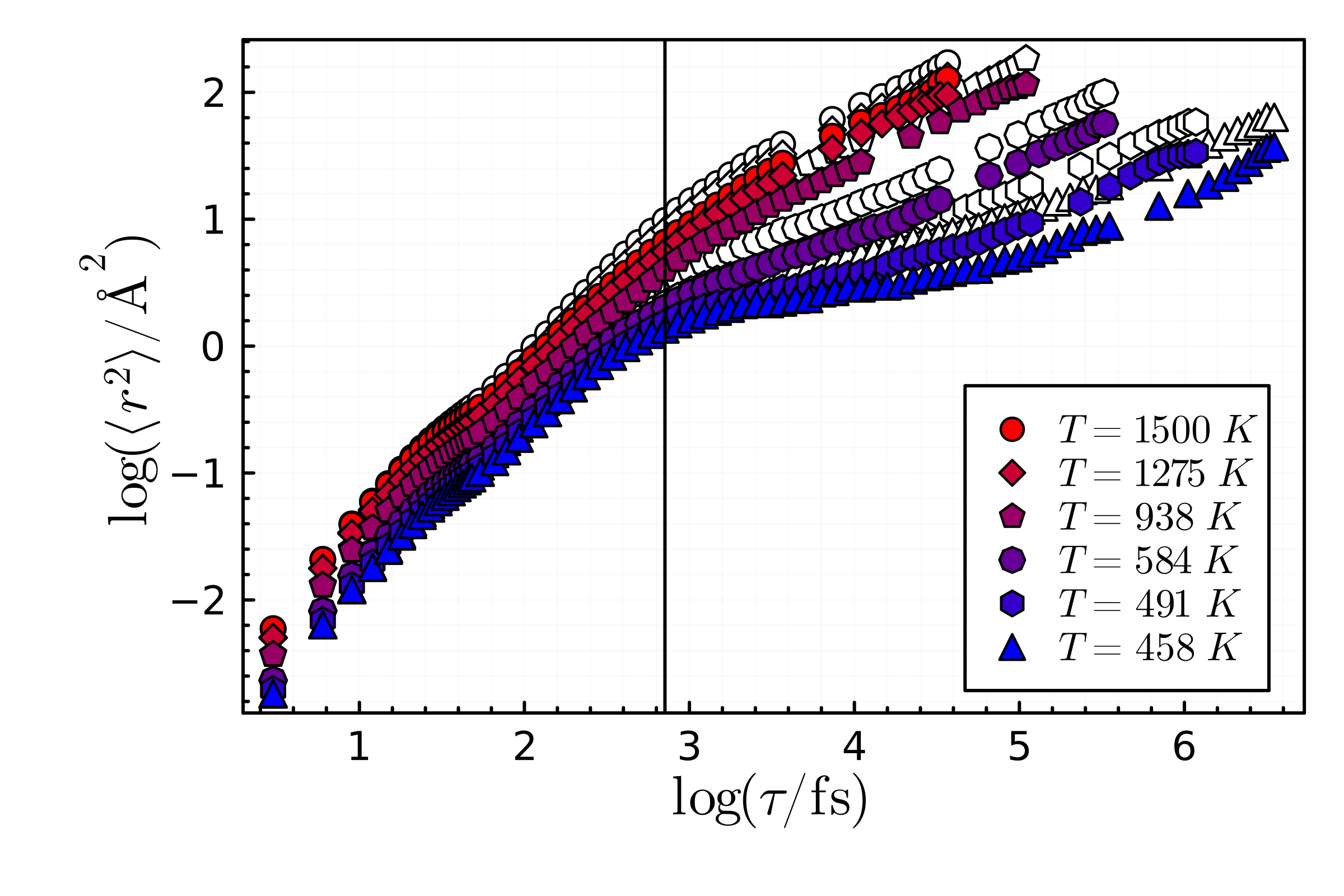}
    \caption{Mean-square displacement data (base 10 logarithm in units \AA$^2$) for select temperatures
    (roughly a tenth of all temperatures simulated) of the all-atom polystyrene de\-ca\-mer ($N=10$);
    the hottest (red circles) and coldest (blue triangles) temperatures simulated are included.
    Chain ends are denoted as open symbols;
    chain middles are denoted as filled symbols.
    The vertical black line shows the timescale chosen for $\langle u^2 \rangle$.
    }
    \label{fig:msd_ps_10}
\end{figure}

\begin{figure}[htbp]
    \centering
    \includegraphics[height=0.3\textheight]{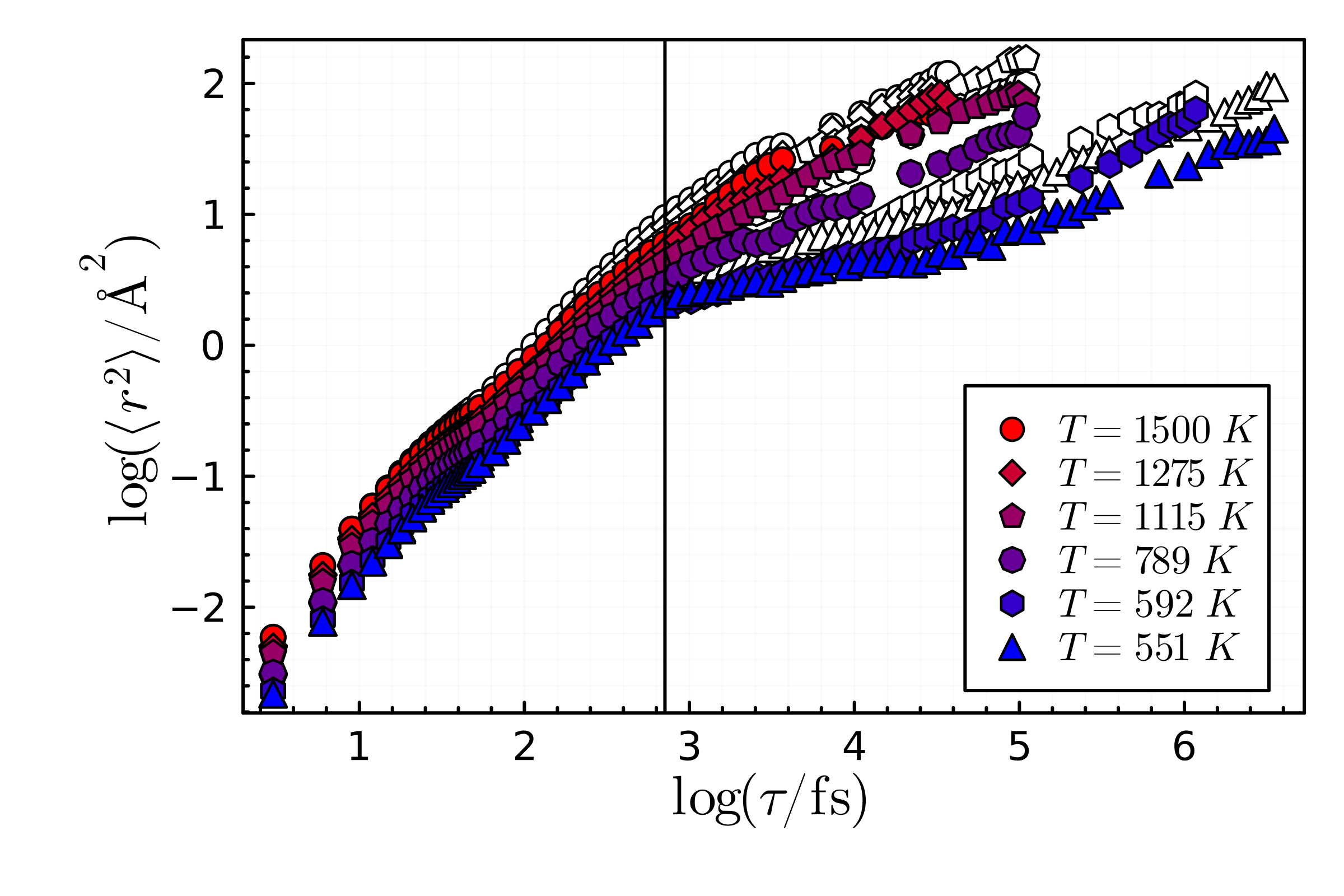}
    \caption{Mean-square displacement data (base 10 logarithm in units \AA$^2$) for select temperatures
    (roughly a tenth of all temperatures simulated) of the all-atom polystyrene he\-ca\-to\-mer ($N=100$);
    the hottest (red circles) and coldest (blue triangles) temperatures simulated are included.
    Chain ends are denoted as open symbols;
    chain middles are denoted as filled symbols.
    The vertical black line shows the timescale chosen for $\langle u^2 \rangle$.
    }
    \label{fig:msd_ps_100}
\end{figure}

As in prior work \cite{hung_universal_2019},
we here define the Debye-Waller factor $\langle u^2 \rangle$ near the cage onset time for the models studied here;
specifically, 
we choose the mean-square displacement at a time delta of 0.275 (or $\approx 10^{-0.55}$) dimensionless time units for both the FJC and FRC and 
0.711 ps (or $\approx 10^{-0.15}$ ps) for AAPS.
These timescale choices are shown below as vertical black lines in Figures \ref{fig:msd_fjc_10}-\ref{fig:msd_ps_100}
(wherein the temperatures are the same as in Figures \ref{fig:isfs_fjc_10}-\ref{fig:isfs_ps_100}).

\subsection*{Dependence of Debye-Waller Factor on Temperature}

Overall (see Figures \ref{fig:dwf_fjc}-\ref{fig:dwf_ps_100} below),
$\langle u^2 \rangle$ is linear with respect to temperature (as expected; see \cite{hung_universal_2019}).
To compare across chain lengths,
we choose a temperature for each model that is at worst 
(for the longest chains, which demonstrate the highest $T_g$ values) 
a slight extrapolation for any given chain length.
We then utilize the line-of-best-fit of $\langle u^2 \rangle$ vs temperature to evaluate $\langle u^2 \rangle$ for that specific temperature for each model: 0.4 for the FJC, 0.6 for the FRC, and 500 K for AAPS.
These resulting values are reported in Figure 3 of the main text.

\begin{figure}[htbp]
    \centering
    \includegraphics[height=0.25\textheight]{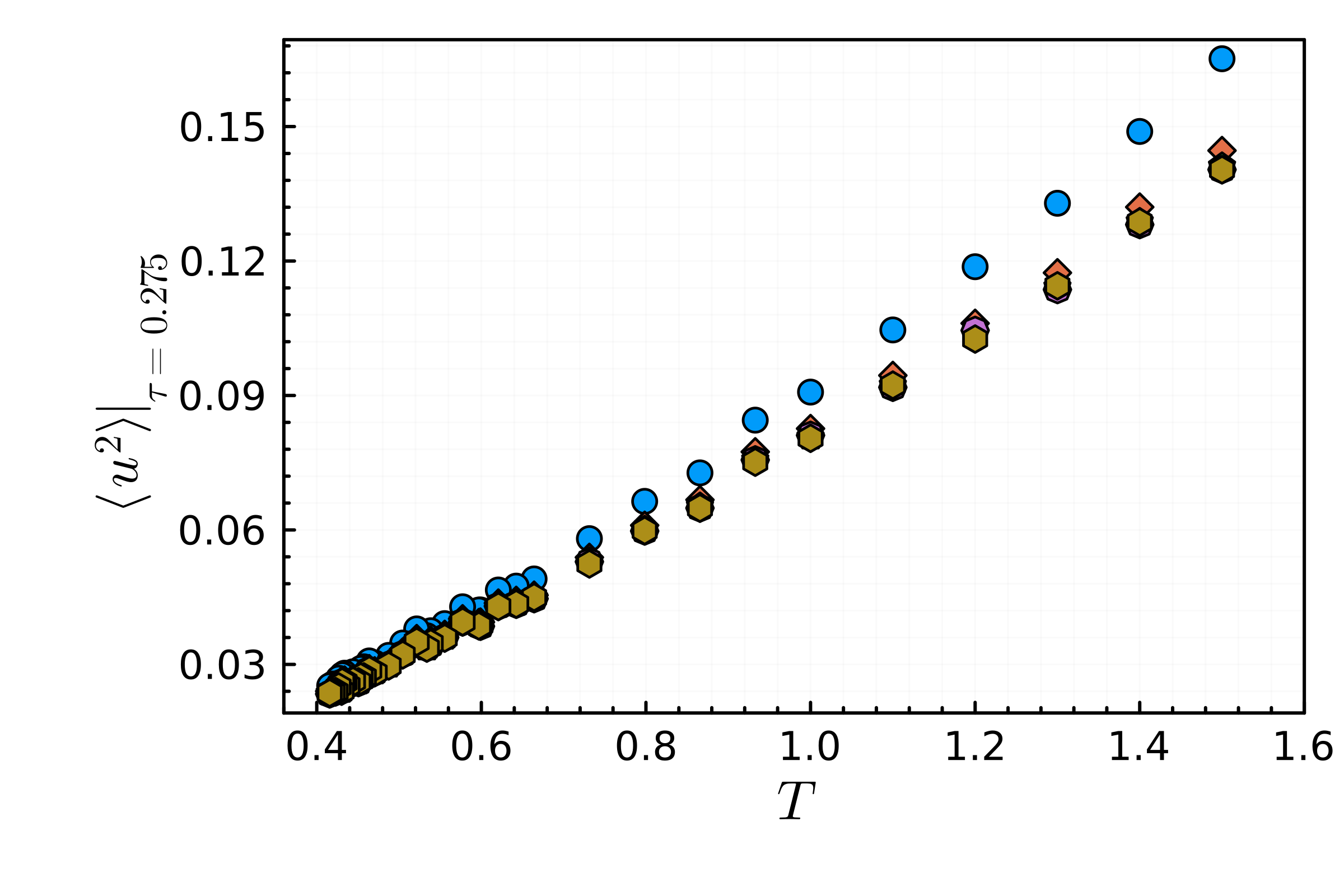}
    \caption{Debye-Waller factor as a function of temperature for the freely-jointed chain of length 10. 
    The chain ends are blue circles ($i = 1$); 
    the pair of monomers bonded to chain ends are orange diamonds ($i = 2$);
    the next pair are green pentagons ($i = 3$);
    the next pair are purple octagons ($i = 4$);
    and the yellow hexagons are the middle pair of segments ($i=5$).}
    \label{fig:dwf_fjc}
\end{figure}

\begin{figure}[htbp]
    \centering
    \includegraphics[height=0.25\textheight]{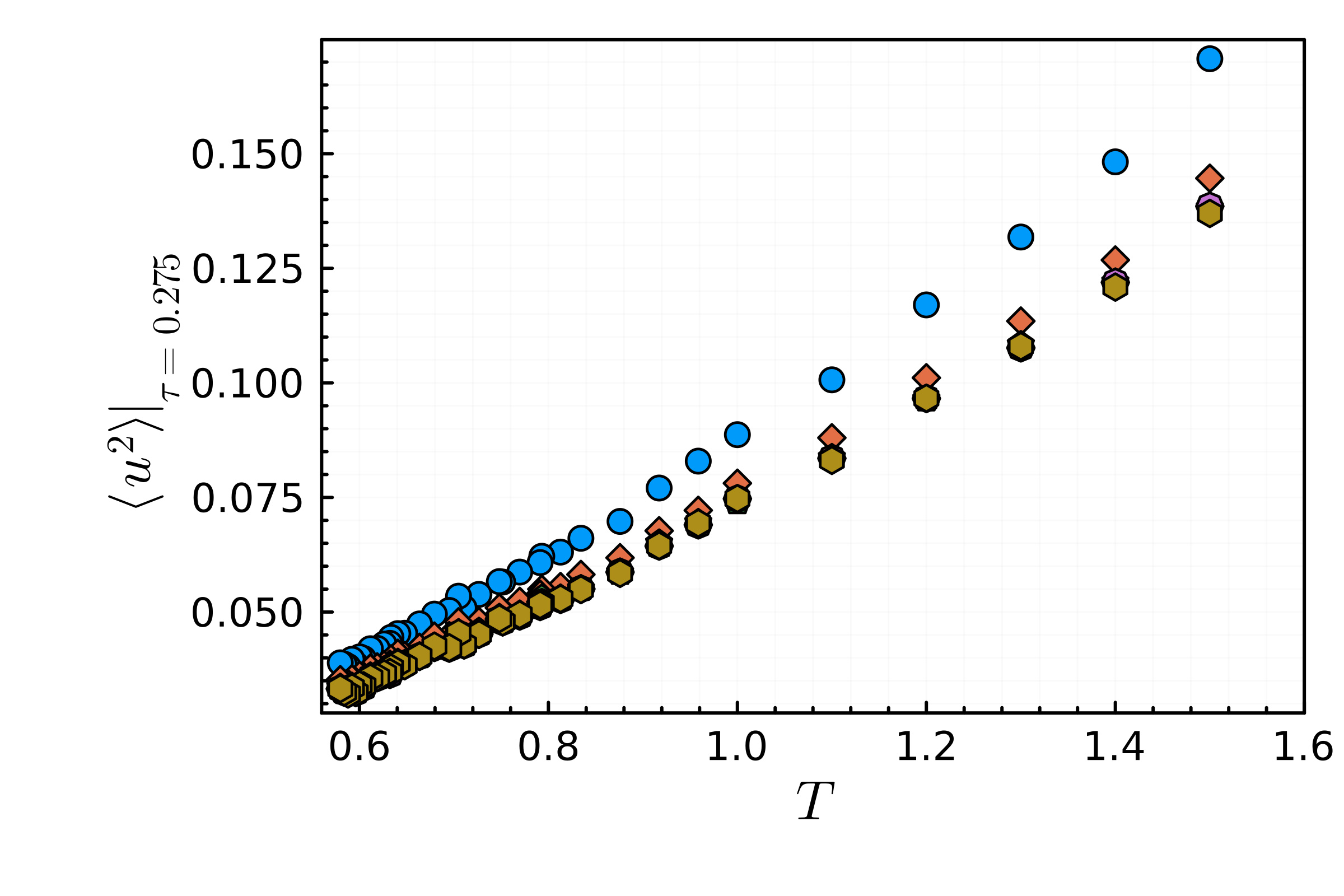}
    \caption{Debye-Waller factor as a function of temperature for the freely-rotating chain of length 10. 
    The chain ends are blue circles ($i = 1$); 
    the pair of monomers bonded to chain ends are orange diamonds ($i = 2$);
    the next pair are green pentagons ($i = 3$);
    the next pair are purple octagons ($i = 4$);
    and the yellow hexagons are the middle pair of segments ($i=5$).}
    \label{fig:dwf_frc}
\end{figure}

\begin{figure}[htbp]
    \centering
    \includegraphics[height=0.25\textheight]{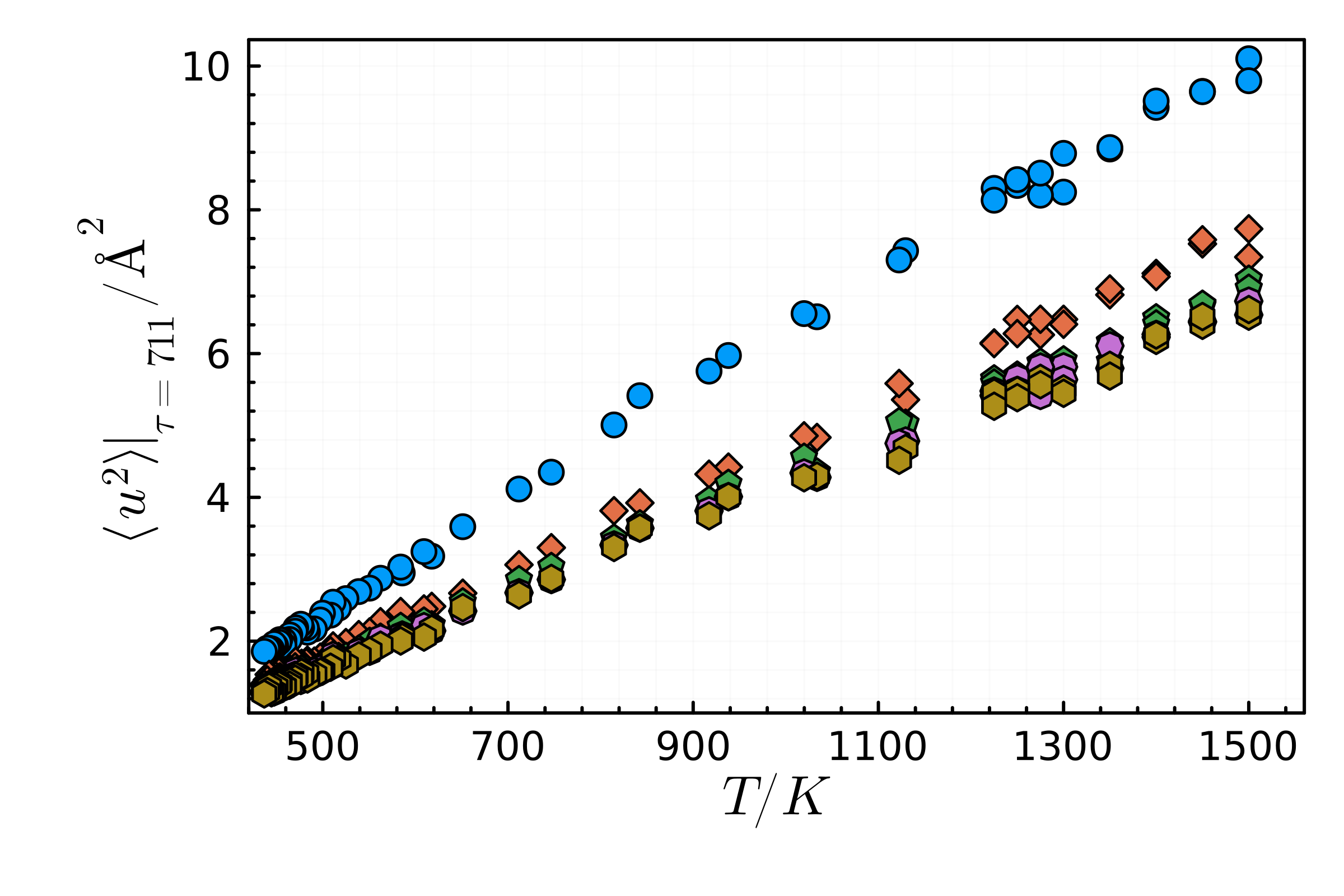}
    \caption{Debye-Waller factor as a function of temperature for the all-atom polystyrene of length 10. 
    The chain ends are blue circles ($i = 1$); 
    the pair of monomers bonded to chain ends are orange diamonds ($i = 2$);
    the next pair are green pentagons ($i = 3$);
    the next pair are purple octagons ($i = 4$);
    and the yellow hexagons are the middle pair of segments ($i=5$).}
    \label{fig:dwf_ps_10}
\end{figure}

\begin{figure}[htbp]
    \centering
    \includegraphics[height=0.25\textheight]{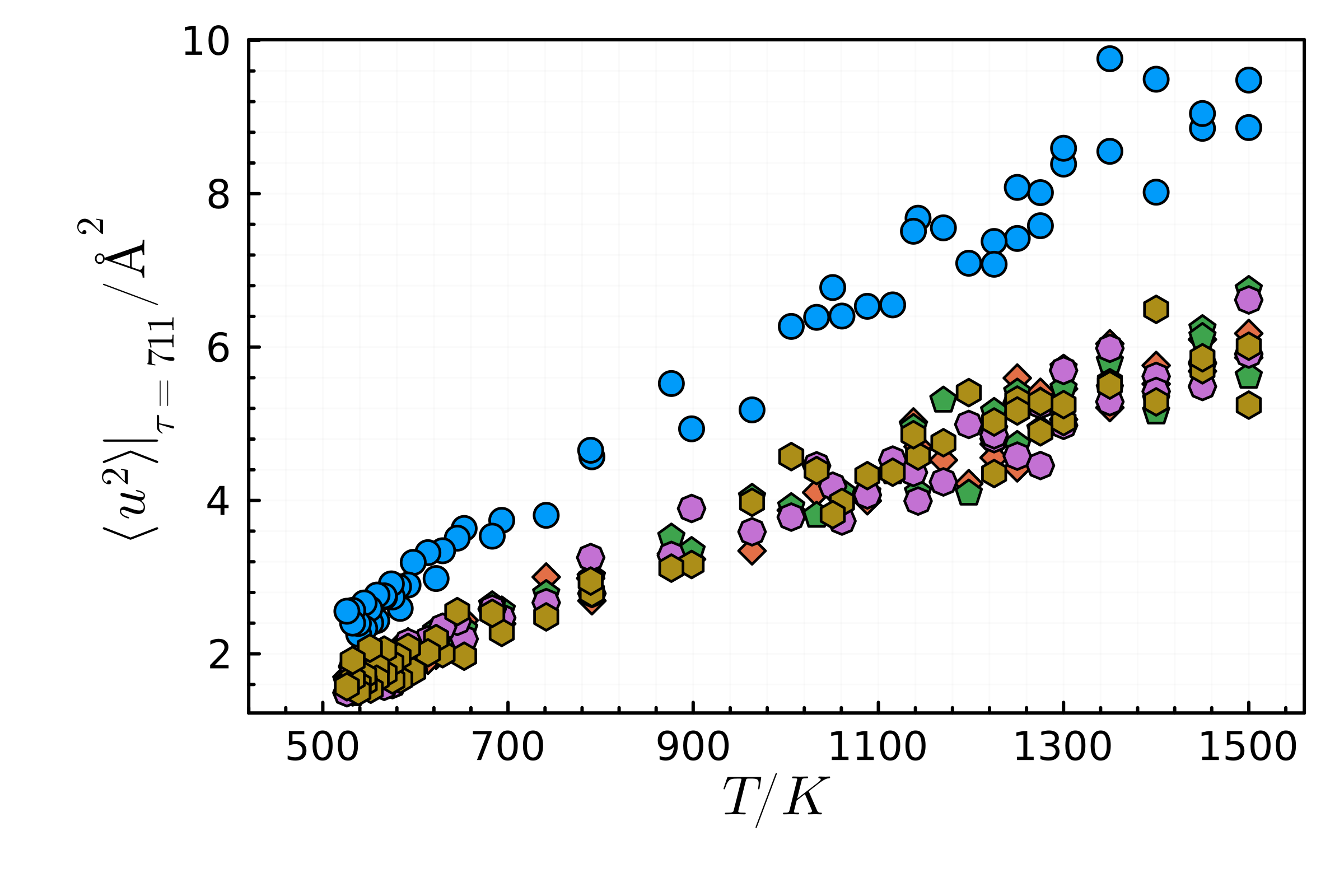}
    \caption{Debye-Waller factor as a function of temperature for the all-atom polystyrene of length 100. 
    The chain ends are blue circles ($i = 1$); 
    the pair of monomers bonded to chain ends are orange diamonds ($i = 2$);
    the next pair are green pentagons ($i = 3$);
    the next pair are purple octagons ($i = 4$);
    and the middle pair of monomers are yellow hexagons ($i = 25$).
    Note that most pairs of monomers ($5 \le i \le 24$) have been omitted for clarity,
    and that noise is increased relative to Figure \ref{fig:dwf_ps_10} due to sampling
    (as monomer count in each AAPS simulation is roughly constant, an order of magnitude increase in chain length corresponds to an order of magnitude less pairs of, e.g., chain ends).
    }
    \label{fig:dwf_ps_100}
\end{figure}

\printbibliography